\documentclass[fleqn,usenatbib]{mnras}

\usepackage{newtxtext,newtxmath}

\usepackage[T1]{fontenc}

\DeclareRobustCommand{\VAN}[3]{#2}
\let\VANthebibliography\thebibliography
\def\thebibliography{\DeclareRobustCommand{\VAN}[3]{##3}\VANthebibliography}

\usepackage{graphicx}	
\usepackage{amsmath}	

\usepackage{lscape}
\usepackage{multirow}
\usepackage{natbib}
\usepackage{diagbox}

\title[Lyapunov Times of TNOs and MBAs]{Numerical analysis of Lyapunov Times for Trans-Neptunian Objects and Main-Belt Asteroids: stability, accuracy, and methodological comparisons}

\author[P. Wajer et al.]{
Paweł Wajer,$^{1}$\thanks{E-mail: wajer@cbk.waw.pl}
Małgorzata Kr\'olikowska,$^{1}$
Jakub Suchecki$^{2,3}$
\\
$^{1}$Centrum Bada\'n Kosmicznych Polskiej Akademii Nauk (CBK PAN) , Bartycka 18A, 00-716 Warszawa, Poland\\
$^{2}$Astronomical Observatory, University of Warsaw, Al. Ujazdowskie 4, 00-478 Warszawa, Poland\\
$^{3}$Cilium Engineering Sp. z o.o., ul. Łokietka 5, 87-100 Toruń, Poland
}

\date{Accepted 2026 February 04. Received 2026 January 16; in original form 2025 October 09}

\pubyear{2015}

\begin{document}
\label{firstpage}
\pagerange{\pageref{firstpage}--\pageref{lastpage}}
\maketitle

\begin{abstract}
We computed Lyapunov times ($T_L$) for a sample of trans-Neptunian objects (TNOs) and outer main-belt asteroids (MBAs) using three numerical approaches: the variational method and two implementations of the renormalization technique. For each object, $T_L$ was derived both from the nominal orbit and from ensembles of 1001 orbital clones, enabling direct comparison between single-orbit and ensemble-based estimates. Across the sample, the methods generally produced consistent results, though larger discrepancies were observed for some MBAs. TNOs, in contrast, displayed greater consistency across methods, likely due to fewer overlapping resonances. Importantly, clone ensembles provided more robust and reliable stability indicators than nominal-orbit computations. Median values from clone populations reduced method-dependent biases and revealed dynamical behaviors that would remain hidden in single-orbit analyses, especially for objects with poorly constrained orbits or evolving in resonant regions. While our study focused on a limited but diverse set of objects, the methodology can be directly extended to larger populations, offering a systematic framework for exploring the long-term stability and dynamical evolution of main-belt asteroids, trans-Neptunian objects or other classes of objects in the Solar System.
\end{abstract}




\section{Introduction}
The Lyapunov time ($T_L$) is one of the most important parameters in celestial mechanics, quantifying the exponential divergence of nearby trajectories in phase space. It serves as a measure of the predictability and stability of an object's motion, with shorter values indicating more chaotic behaviour. In the context of asteroid dynamics, the Lyapunov time is particularly useful for identifying chaotic regions in the Solar System, such as those associated with mean-motion or three-body resonances (e.g., asteroid-Jupiter-Saturn interactions). Moreover, $T_L$ is closely linked to an asteroid's dynamical lifetime ($T_D$), which describes the timescale over which an object's orbit evolves to the point of crossing planetary orbits, ultimately leading to ejection or collision. The relation between $T_L$ and $T_D$ is given by \citep{Lecaretal1992, LevisonandDuncan1993, Murisonetal1994}:
\begin{equation}
\frac{T_D}{T_0}\approx a\left(\frac{T_L}{T_0}\right)^b ,
\end{equation} 
where $a$ and $b$ are positive constants, and $T_0$ is an appropriately normalized period of the asteroid. However, \citet{MilaniandNobili1992} gave an example of object for which this simple relation does no hold. They found that the asteroid (522) Helga has relatively short  Lyapunov time of 6900\,yr but its orbit does not change significantly for 1000$T_L$. Helga is an example of objects moving on stable chaotic orbits. Stable chaotic orbits are characterized by a short Lyapunov time, indicating rapid local divergence of nearby trajectories, yet their proper elements exhibit only limited variations over millions of years, resulting in long-term dynamical stability despite underlying chaos. For such objects relationship between $T_L$ and $T_D$ is exponential \citep{Morbidelli1996}:
\begin{equation}
T_D\approx ce^{T_L/T_0}  .
\end{equation}
\citet{Winteretal2010} used simply modification of Lyapunov time to identity of objects which could move in the stable chaotic orbits or be escapers (i.e., objects that moving on chaotic trajectories and, due to substantial variations in their orbital elements, ultimately escape their initial dynamical region, typically by crossing planetary orbits or being ejected from the system).

The computation of the Lyapunov time in practice is based on numerical methods, as analytical solutions are generally not feasible for complex dynamical systems. Each numerical method has its advantages and limitations \citep{Tancredietal2001}, and the choice of method depends on various factors such as the specific application and available computational resources. There are two main numerical approaches commonly used. The variational method utilizes the evolution of an initially small deviation vector between two nearby trajectories, integrating variational equations alongside the object's equations of motion. While widely used, this approach requires precise numerical differentiation and may be sensitive to integration errors. Alternatively, the neighbour trajectories method (renormalization method or two paricle method) periodically rescales the deviation vector to avoid numerical underflow, providing a more robust estimate of the Lyapunov time. This technique ensures long-term stability of the calculations but may introduce incorrect estimate of $T_L$ in certain cases \citep{HolmanandMurray1996, Tancredietal2001}. The choice of method, along with factors such as the adopted Solar System model (e.g., number of included planets) and the accuracy of initial orbital elements, significantly affects the precision of $T_L$ estimates (see Section \ref{Sect:Results} where more detailed discussion on the influence of these factors is presented). 

Over the past decades, Lyapunov time have been calculated for numerous asteroids across a wide range of studies, often as part of broader efforts to characterize orbital stability or identify chaotic regions in the Solar System. While many individual publications report such values, two sources stand out as the most comprehensive: the Asteroids Dynamic Site (AstDyS-2) webpage \citep{KnezevicandMilani2012}, and the Asteroid Families Portal (AFP) \citep{NovakovicandRadovic2019}, both of which provide Lyapunov times for large numbers of objects. In addition to these databases, several other studies have presented Lyapunov times for smaller samples of asteroids. A more detailed overview of these works, along with relevant references, is provided in Section~\ref{Sect:Results}.

In these two databases AstDyS-2 and AFP, as well as in the majority of published studies, the Lyapunov time is typically computed for a single, nominal best-fit orbit. This approach does not take into account the uncertainties in the initial orbital parameters or the potential spread of possible trajectories represented by cloned orbits. Consequently, the resulting Lyapunov times usually reflect the dynamical behaviour of only the best-fit orbit, potentially overlooking the range of dynamical outcomes that may arise due to the uncertainty of the orbital elements. 

In this work, we analyze the above-mentioned methods for computing the Lyapunov time  to a diverse sample of small bodies located in different regions of the Solar System. Our dataset includes objects from the trans-Neptunian region (TNOs), the Scattered Disc and the Main Belt, each representing different dynamical environments and stability regimes. A second key objective of this article is to demonstrate the advantages of the Lyapunov time calculations based on ensembles of orbital clones compared to estimates derived from a single nominal orbit. In addition, we discuss the problem of Lyapunov time convergence and its impact on the predictability of the stability of the examined orbits.

By analyzing these asteroids and distant minor planets, we aim to assess how different numerical approaches to the Lyapunov time estimation perform under varying orbital conditions, particularly in regions affected by planetary perturbations and resonances.

This article is organized as follows. In Sect.~\ref{Sect:Methods}, we provide in detail the numerical methods used to calculate the Lyapunov time, namely the variational method and the method of nearby trajectories. We also introduce and define the key parameters and notations that are used consistently throughout the paper. Sect.~\ref{Sect:Data} presents the general information of the selected celestial objects analyzed in this study, together with their initial orbital elements. Sect.~\ref{Sect:Results} is devoted to the presentation and interpretation of the results obtained from our numerical simulations, where we also discuss their implications for the dynamical stability of the studied orbits. Sect.~\ref{Sect:TNO_oMBA_Accuracy} addresses the relationship beetwen the orbital accuracy and its effect on the estimation of Lyapunov times. Next (Sect.~\ref{Sect:TNO_oMBA_Comparison}), we examine the discrepancies in Lyapunov time calculations for trans-Neptunian objects (TNOs) and main-belt asteroids (MBAs) across various numerical approaches, with a particular focus on the consistency between results based on nominal orbits and those derived from virtual-asteroid (VA) distributions. Finally, we summarize our main findings and provide concluding remarks in Sect.~ \ref{Sect:Summary}.

\section{Methods}
\label{Sect:Methods}
In our analysis, the dynamical evolution of the studied objects was modelled by integrating a nine-body problem in a barycentric reference frame, with the Sun as the central mass and seven planets from Venus to Neptune included explicitly. The mass of Mercury was added to that of the Sun, and the test particle, representing a VA, was treated as massless. Mercury was not modelled as a separate body due to its relatively small mass and short orbital period, which have a negligible long-term perturbative effect on the dynamical evolution of the analysed objects, while significantly increasing the computational cost. All calculations were performed using the open-source N-body integrator REBOUND \citep{ReinandLiu2012}, with the high-accuracy IAS15 integration scheme \citep{ReinandSpiegel2015}, which is a 15th-order adaptive integrator based on the Gauss–Radau method \citep{Everhart1985}. IAS15 is optimized for long-term simulations of planetary systems and ensures numerical stability and precision over extended time spans, making it suitable for the computation of dynamical indicators such as the Lyapunov time.

To estimate the Lyapunov times ($T_L$) of the selected asteroids, two complementary approaches were employed: one based on the integration of variational equations, and the other based on the evolution of initially neighboring trajectories. For each method, we outline the implementation details, their adaptation to our problem, and discuss their advantages and limitations.

\subsection{Lyapunov time calculation - variational equations}
In the variational approach, the Lyapunov exponent is computed by integrating the system of linearized equations of motion, also called tangent equations, which track the evolution of an infinitesimal deviation vector along a reference trajectory. This procedure tracks the evolution of an infinitesimal deviation vector, from which the exponent can be derived. For an $n$-body system, the variational equations for the $k$-th body take the form: 
$$
\ddot{\vec{\xi}}_k = \sum_{\substack{i = 1 \\ i \neq k}}^n \left( \frac{Gm_i}{r_{ik}^3} \vec{\xi}_{k} - 3 \frac{Gm_i \vec{r}_{ik}}{r_{ik}^5} (\vec{r}_{ik} \circ  \vec{\xi}_{k}) \right) .
$$

where $m_i$ is the mass of the $i$-th body, $\vec{r}_i$ and $\vec{r}_k$ are the position vectors of bodies $i$ and $k$, respectively, and $\vec{r}_{ik} = \vec{r}_i - \vec{r}_k$. The derivation of the variational (tangent) equations used here is presented in detail in Section 3.1 of \citet{ReinandTamayo2016}.

The instantaneous growth rate of the deviation vector is measured by the Lyapunov characteristic indicator,
\begin{equation}
\lambda(t) = \frac{1}{t} \ln \frac{\xi_k(t)}{\xi_k(0)} ,
\end{equation}
from which the Lyapunov exponent is obtained as
\begin{equation}
\lambda = \lim_{t \to \infty} \lambda(t) .
\end{equation}

The Lyapunov time, $T_L$, is defined as the inverse of $\lambda$. For finite integration times, we use the finite-time Lyapunov indicator $T_L(t) = 1/\lambda(t)$, which is reported in Sect.~\ref{Sect:Results}.

Although the equations of motion in the $n$-body problem are nonlinear, the associated variational equations are linear with respect to the deviation vector $\vec{\xi}$. This property allows the initial norm of $\vec{\xi}(0)$ to be chosen arbitrarily, since only the relative growth rate of the perturbation matters for the computation of the Lyapunov exponent. However, if necessary to keep the upper range within the machine’s numerical limits, a rescaling procedure can be applied. In our implementation, a fixed numerical threshold ($10^8$) was used to renormalize the deviation vector when necessary, and the same criterion was applied uniformly to all objects.

\subsection{Lyapunov time calculation - neighbour trajectories}
An alternative to the variational approach for computing the Lyapunov exponent is the neighbour trajectories method, originally proposed by \citet{Benettinetal1976}. Instead of solving the variational equations, this method estimates the divergence rate, by numerically integrating a nominal trajectory and a nearby test particle initially separated by a small displacement. The distance between these two trajectories is monitored over time, and its exponential growth is used to compute the Lyapunov exponent. Let $d(0)$ denote the initial separation between the two trajectories, and $d(t)$ the separation at time $t$. The finite Lyapunov characteristic indicator $\lambda(t)$ is given by:
\begin{equation}
\lambda(t)=\frac{1}{t}\ln  \frac{d(t)}{d(0)} .
\end{equation}

The mentioned above exponential growth of distance between trajectories is achievable only for a short term so it is needed to perform periodical renormalizations. After fixed time intervals $\Delta t$, the separation $d(t)$ is rescaled to its original length $d(0)$, and the growth factor is recorded at each step. The Lyapunov characteristic indicator is then approximated as:

\begin{equation}
\label{eqle2}
\lambda(t) \approx \frac{1}{N\Delta t} \sum_{n=1}^{N} \ln \frac{d(t_n)}{d(0)} ,
\end{equation}
where $d(t_n)$ is the separation between the trajectories at the end of the $n$-th interval (before renormalization), and $N \Delta t$ is the total integration time, $t$. Finally, the Lyapunov exponent is defined as the limit of the finite time estimate as $N\rightarrow \infty$.

The function $d(t)$, representing the distance between two nearby trajectories, is constrained from above by theoretical considerations (i.e., it should not exceed the regime of linear approximation) and from below by the limits of numerical precision. In practice, for typical objects in the Solar System, the optimal range for $d(0)$  is $10^{-8}$ to $10^{-4}$ au and $d(t) \le 10^{-4}$ au \citep{Murison1995}. Two values for $d(0)$ were chosen for our calculations: $10^{-7}$ au and $10^{-6}$ au. 

The neighbor trajectories method does not require explicit computation of variational equations and is therefore particularly useful when such equations are either unavailable or computationally expensive to solve. For this reason, the method has been widely adopted in practice, likely due to its simplicity and the lack of built-in variational solvers in most popular n-body integration software.

However, despite its practical usefulness, the method is known to exhibit serious anomalies in certain cases, most notably, the Lyapunov exponent values tend to be overestimated and reach saturation too rapidly. The first attempt to explain this phenomenon was made by \citet{HolmanandMurray1996}, who suggested that the issue may have a hidden mathematical origin related to the method itself (i.e. the rescaling procedure). A few years later, this hypothesis was challenged by \citet{Tancredietal2001}, who presented counterarguments, particularly the absence of similar problems in the case of rescaling of variational equations distance. As noted by \citet{Tancredietal2001}, the neighbour trajectories method is not recommended in regions of regular or near-regular motion, where the variational approach yields more reliable Lyapunov exponent estimates. In strongly chaotic regimes, the neighbour trajectories method can still produce acceptable results, but care must be taken: multiple tests with varying initial separations and renormalization intervals are advised to ensure the stability of the final value.

\section{The sample of small bodies and starting orbits for the dynamical study}\label{Sect:Data}
For the long-term dynamical study, we selected twelve representatives from two dynamically distinct groups of small solar system bodies\footnote{we follow the JPL Small Body Database classification, see \tt https://ssd.jpl.nasa.gov/tools/}. These are:

\begin{enumerate}
	\item three Trans-Neptunian Objects (TNOs) with semi-major axes 39\,au $< a <$ 41\,au and eccentricities $e < 0.18$: 2010 EL$_{139}$, 2010 JK$_{124}$, and (471165) 2010 HE$_{79}$; these are close to a 2:3 mean-motion resonance with Neptune as shown in Fig.\ref{fig:TNO_classes},
	\item four other TNOs (2010 KZ$_{39}$, 2010 FX$_{86}$, (471143) Dziewanna, and 2010 JJ$_{124}$) with perihelion distances between 23\,au and 44\,au covering a wide range of eccentricies between 0.05 and 0.73 (semi-major axes between 45\,au and 86\,au). Two of them are often referred as objects of Scattered Disc (SDOs); see Fig.\ref{fig:TNO_classes},
	\item five outer Main Belt Asteroids (oMBA: El Leoncito, Omarkhayyam, Oda, Helga, and Devota) with 3.48\,au $< a <$ 3.76\,au, 3.03\,au  $< q <$ 3.48\,au, and 0.04 $< e <$ 0.14; see Sect.~\ref{Sect:oMBA}.
\end{enumerate}

All seven selected TNOs, listed in Table~1 of \cite{Sheppardetal2011}, were discovered in 2010 during the OGLE (Optical Gravitational Lensing Experiment) Carnegie Kuiper Belt Survey \citep{Udalskietal2015}, which contributed to the discovery of distant Solar System bodies. They were chosen for this study because, at the time our integrations were initiated, their individual dynamics had been only sparsely investigated. Radius estimates suggest sizes in the range $\sim 160$--600\,km \citep{Sheppardetal2011}. 

In contrast, oMBA objects were selected among those with observational arcs spanning more than 30 oppositions, ensuring that their Lyapunov times had already been determined in the literature. A comparative discussion of our results with previous determinations is given in Sect.~\ref{Sect:oMBA}.  

General orbital characteristics and data arcs used for orbit determination of the twelve objects studied are summarized in Table~\ref{t7}. Orbital parameters were taken from the IAU Minor Planet Center\footnote{https://www.minorplanetcenter.net/db\_search/} (MPC) in October 2022 for TNOs and in March 2023 for oMBAs. At the same time, we retrieved the positional observations employed in our orbit determinations. As of August 2025, data arcs have been extended for most of the sample (for three TNOs: 2010 KZ$_{39}$, 2010 FX$_{86}$, and 2010 JK$_{124}$ they are up to date). These objects are treated here as representative examples illustrating how uncertainties in orbit determination, the use of nominal orbits, and methodological choices affect Lyapunov time estimates. 

The methods used for osculating orbit determination are described in \cite{Krolikowskaetal2009} and references therein. Our results are in excellent agreement with those of the MPC and JPL Small Bodies Database \citep{KrolikowskaandDones2023}. In our dynamical model, perturbations from all eight planets and Pluto are included, making it somewhat simpler than the MPC or JPL models. This simplification is intentional: the derived orbits serve as initial conditions for the long-term integrations presented here, which employ a nearly identical perturbation model. This ensures maximum internal consistency without neglecting relevant dynamical effects.  

The resulting orbital elements are given in Table~\ref{tab:starting-orbits}. A comparison of the uncertainties in the orbital element determinations (with the semimajor axis being particularly well suited to assess orbit quality) shows that all oMBA orbits have very small uncertainties, of order $\delta a/a = $ 1--4$\times 10^{-9}$, where  $\delta a$ denotes the uncertainty in $a$. In contrast, the TNOs discovered in 2010 exhibit significantly larger uncertainties, with $\delta a/a$ ranging from 0.6 to 23.1 in units of $10^{-5}$, i.e. typically about four orders of magnitude greater, and additionally show a much broader spread.

Next, we constructed swarms of 1001 virtual asteroids (hereafter VAs) based on the obtained nominal orbits (that is: the nominal orbit + 1000\,VAs), following the approach introduced by \citet{Sitarski1998}; see also, for example, \cite{KrolikowskaandDybczynski2020} and references therein. This VA method employs Monte Carlo–generated swarms of orbital elements to propagate observational uncertainties in dynamical studies. This step is crucial, as it provides the basis for assessing how orbital uncertainties affect the Lyapunov time estimates analyzed in the following section.

\section{Results}
\label{Sect:Results}
As outlined in the previous section, for each of the 12 selected objects (7 trans-Neptunian objects  and 5 outer main-belt asteroids) we computed the Lyapunov time ($T_L$) using three independent methods, applied to 1001 virtual clones. This approach enables a comprehensive assessment of dynamical stability, both through statistical analysis of the entire clone population and by comparative study  of the nominal orbit alone. 

This section is divided into three subsection. In the first (Sect.\ref{Sect:Stat}), we describe the statistical parameters and the methodology used to analyze the  results. The following two subsections present the outcomes, separately for the two dynamical populations studied: TNOs (Sect.\ref{Sect:TNO}) and oMBA (Sect.\ref{Sect:oMBA}). 

\subsection{Statistical methods and parameters}
\label{Sect:Stat}
The calculation of the Lyapunov time for asteroids and other small bodies in the Solar System is a complex task that depends on multiple numerical and physical factors. Among the most important are the initial orbital conditions, the numerical integration scheme, and the adopted Solar System model. In particular, the model's configurations, especially the choice of perturbing planets, can significantly affect the computed orbital evolution and, consequently, the estimated $T_L$. Simplified models that, for example, include only the giant planets may yield substantially different results compared to more complete configurations that also account for the terrestrial planets, Pluto, or even massive asteroids. The inclusion or omission of specific gravitational perturbers directly influences the calculated dynamical time-scales and the identification of chaotic versus regular orbital behaviour (see \ref{Sect:oMBA}, where this issue is examined in greater detail). 

Another critical factor is the uncertainty in the orbital elements. Because many Solar System bodies exhibit chaotic dynamical behaviour, even small observational errors in parameters such as semi-major axis, eccentricity, or inclination can lead to notable differences in the estimated $T_L$, resulting in divergent stability predictions. \citep{Murisonetal1994} showed that the computed Lyapunov time ($T_L$) for chaotic asteroids such as (522) Helga can vary significantly with small changes in initial conditions, particularly near mean-motion resonances. Considering these aspects, estimating the $T_L$ (and its associated uncertainty) from a large number of VAs allows for a more accurate assessment of this parameter for a given body.
	
In our study, we computed $T_L$, using three independent methods, for each object within a swarm of 1001\,VAs, constructed as described in Sect.~\ref{Sect:Data}. In this section, especially in Tables \ref{tab:TNO-LT} and \ref{tab:MBA-LT}, symbol 'V' refers to the Lyapunov time estimated using the variational method, 'N1' and 'N2' refer to two variants of the renormalization method. In both these variants, N1 and N2, the separation $d(t)$ between nearby trajectories is kept below $10^{-4}$ au. For N1, the initial separation is fixed at $d(0) = 10^{-6}$ au. For N2, $d(0) = 10^{-7}$ au.

From the resulting values of $T_L$, we calculated statistical parameters, such as the median ($Med$), the first ($Q_1$) and third ($Q_3$) quartiles, and the interquartile range, defined as $IQR = Q_3 - Q_1$, to characterize the spread of $T_L$ values. We also calculated their skewness ($A_Q$) defined as: 
\begin{equation}
A_Q=\frac{Q_1+Q_3-2Med}{IQR} .
\end{equation}

A value of $A_Q$ equal to zero indicates a symmetric distribution, while negative values correspond to a left-skewed distribution with an elongated left tail. Positive values indicate a right-skewed distribution with an elongated right tail.

To check the consistency of the $T_L$ estimates obtained using different numerical methods, we calculated the relative spread of median $T_L$ values and the relative spread of $T_L$ for the nominal orbit. The first parameter is defined as:
\begin{equation}
\delta_{Med}=\frac{\max{(Med)}-\min{(Med)}}{\overline{Med}} ,
\end{equation}
where $\max{(Med)}$ and $\min{(Med)}$ are the largest and the smallest values of $Med$ among the three median values calculated with different methods. In a similar way, we define the second parameter, denoted as $\delta_{nom}$. 

In addition to this statistical characterization, we examined the temporal evolution of the Lyapunov time, $T_L(t)$, for all clones. This approach offers further insight into the dynamic nature of an object's orbit, going beyond Lyapunov time estimates derived solely from the  trajectory of a single VA, particularly the nominal orbit. Based on the evolution patterns of $T_L(t)$, the analysed orbits can be categorized into three classes:
\begin{enumerate}
\item
\textit{Stable orbits}: For all VAs, $T_L(t)$ exhibits a long-term increasing trend over the integration interval, without converging to a constant value. This indicates either divergence toward infinity or a very large $T_L$, suggesting a likely stable orbit for which determining the exact Lyapunov time is problematic. In such cases, we estimate a lower bound for the Lyapunov time of a given clone by taking its value at the end of the integration period, $T_L(t_0)$, where $t_0$ denotes the integration time. 

\item
\textit{Unstable orbits}: For all VAs, $T_L(t)$ converges to a constant value over the integration time. This behaviour indicates a likely unstable orbit, as the convergence of $T_L(t)$ to a finite asymptotic value reflects the exponential divergence of nearby trajectories. In these cases, the Lyapunov time of a given VA can be reliably estimated as the limiting value of $T_L(t)$ reached at the end of the integration. 

\item
\textit{Unresolved orbits}: In this category, some VAs show convergence of $T_L(t)$ to a finite value, while others do not. This inconsistency across the clone ensemble prevents a reliable Lyapunov time from being assigned to the object. In such cases, we take the value $T_L(t_0)$ at the end of the integration interval as a provisional estimate of Lyapunov time, keeping in mind that this value may not reflect the true value of $T_L$ of the object.
\end{enumerate}

An important consideration in computing of $T_L$ estimates is the choice of integration timespan, which must be sufficiently long to ensure convergence (see, e.g., \citealt{PerezHernandezandBenet2019} and references therein). In our study, we performed orbital simulations over a 15-million-year period, which was generally adequate to obtain $T_L$ estimates for the analysed objects. However, in a few cases even this integration interval may not suffice to yield a reliable $T_L$ value. The following sections provide a more detailed discussion of the two dynamical groups, with each object classified into stable, unstable, or unresolved orbital classes.

\subsection{Trans-Neptunian objects}
\label{Sect:TNO}
As noted in Sect.~\ref{Sect:Data}, all of the selected TNOs were discovered in 2010 during the OGLE project. The characteristics of their orbital elements, as well as the lengths of their observational arcs, are summarized in Table~\ref{t7} (see Appendix Sect.~\ref{sec:orbital-elements}). The orbital elements derived from these data arcs, which then were used as the starting nominal orbits for our dynamical studies are listed in Table \ref{tab:starting-orbits}. Column~[10] shows the quantity $\delta a /a$ -- a good measure of orbital quality.

According to these estimates, the orbits of the two numbered objects in the sample, (471143) Dziewanna and (471165) 2010 HE$_{79}$, along with 2010 JK$_{124}$, are the most precisely determined within TNO sample, owing to their relatively long observational arcs and the availability of pre-discovery images. The poorest quality orbits belong to 2010 JJ$_{124}$ and 2010 JK$_{124}$, which have the shortest data arc.

For three TNOs analyzed in this work, 2010 FX$_{86}$, 2010 EL$_{139}$ and 2010 JK$_{124}$, the $T_L$ (in fact the Lyapunov exponent, which is the inverse of the Lyapunov time) is also available in the AstDyS-2 online database.\footnote{The values reported here correspond to the state of the AstDyS-2 database as of January~12,~2026.} These values have been compiled and included in the last column of Tab.~\ref{tab:TNO-LT}.

\subsubsection{Overview of Orbital Characteristics of the Analysed TNOs}
\label{Sect:TNO_Intr}
\begin{figure*}
	\centering
	\includegraphics[width=18.00cm]{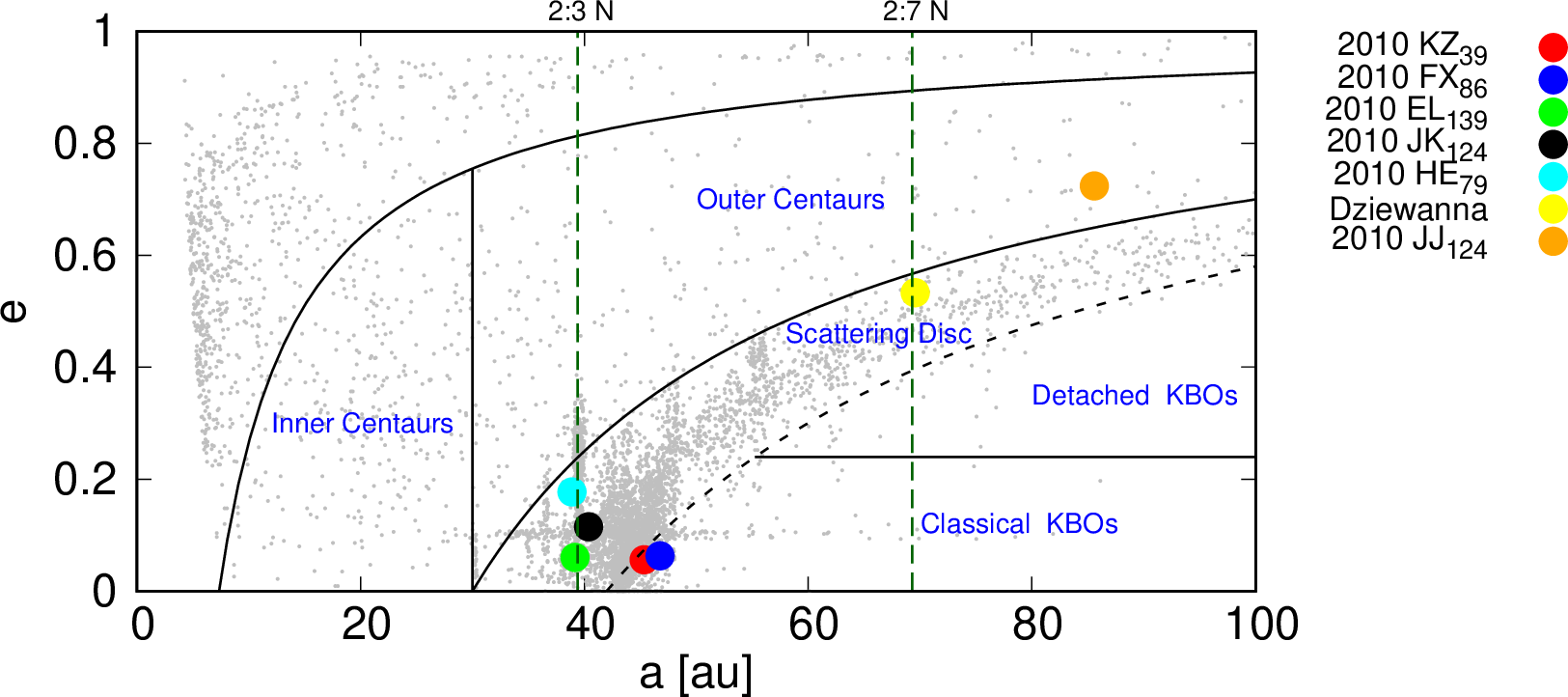}
	\caption{Distribution of the analysed TNOs in the $a-e$ plane, with the main dynamical classes indicated: Centaurs, scattered disc, classical, detached, and resonant (Plutinos). Grey points illustrate the overall distribution of the TNO population, based on data retrieved from the Minor Planet Center webpage on 14 January 2026.}
		\label{fig:TNO_classes}
\end{figure*}

The analyzed TNOs exhibit a wide range of orbital parameters, as shown in Fig.~\ref{fig:TNO_classes}. In particular, their eccentricities range from 0.057 to 0.724, indicating the presence of both nearly circular and highly elongated orbits. The broad distribution of semi-major axes (from 38.9\,au to 85.6\,au) and inclinations (from 16\degr -- 38 \degr) further highlights the dynamical diversity of the sample, which includes representatives of several distinct trans-Neptunian populations.

To provide context for their classification, we adopt definitions based on \citet{Gladmanetal2008} with further modifications from \citet{Khainetal2020}:

\begin{itemize}
 \item \textbf{Centaurs:} objects with perihelion distances $7.35\,\mathrm{au} < q < 30\,\mathrm{au}$; 
          those with semi-major axes $a<30\,\mathrm{au}$ are \textit{inner Centaurs}.

 \item
\textbf{Scattered disc objects (SDOs):} approximately defined by $30\,\mathrm{au} < q \lesssim 42\,\mathrm{au}$, with most objects having perihelia near Neptune ($q \sim 30$\,au). The upper limit is not strict and serves as a guideline for classification. Due to gravitational interactions -- primarily with Neptune -- the orbits of SDOs can evolve over time, potentially approaching Neptune's orbit and increasing the likelihood of further scattering or even ejection from the Solar System.

 \item \textbf{Classical population:} objects with $q\gtrsim 42\,\mathrm{au}$ and eccentricity $e<0.24$.
 
 \item \textbf{Detached population:} objects with $q\gtrsim 42\,\mathrm{au}$ and eccentricity $e>0.24$.

 \item \textbf{Resonant TNOs:} objects trapped in mean-motion resonances with Neptune; for example, \textbf{Plutinos} occupy the 2:3 resonance at $a\simeq 39.4\,\mathrm{au}$. 
\end{itemize}

Additionally, TNOs with higher orbital inclinations and/or eccentricities are collectively reffered as \textbf{dynamically excited TNOs}. This group includes hot classical TNOs (characterized by higher inclinations and relatively low eccentricities) as well as resonant, scattering, and detached classical TNOs \citep{Pikeetal2021}.

Applying these definitions to our sample, the dynamical diversity of the analyzed TNOs is evident. Classical TNOs include 2010 KZ$_{39}$ and 2010 FX$_{86}$, both belonging to the Hot Classical subpopulation, while 2010 JJ$_{124}$ falls within the perihelion range defined for Centaurs. The objects 2010 EL$_{139}$ and (471165) 2010 HE$_{79}$ lie near the 2:3 mean-motion resonance with Neptune. Our analysis {\bf confirms} that 2010 EL$_{139}$ is indeed in this resonance, classifying it as a representative of the Plutino population (see Sect.~\ref{Sect:TNO_Unresolved} for details). In turn, (471165) 2010 HE$_{79}$ is particularly interesting due to its binary nature and is classified as an excited TNO \citep{Pikeetal2021}. (471143) Dziewanna moves in the 2:7 resonance with Neptune; although it is not firmly trapped in this resonance \citep{Munoz-Gutierrezetal2025}, its motion within the resonance allows it to be also classified as a resonant object. 
		
The case of 2010 JK$_{124}$ is especially noteworthy: although its semi-major axis places it near the 2:3 mean-motion resonance with Neptune, it is not currently locked in this configuration. Our calculations indicate, however, that it occasionally enters the much weaker 9:14 resonance with Neptune\footnote{The identification of the mean motion resonances was performed by analyzing the time evolution of the principal resonance angle, defined by \citet{Gallardo2006} as $\sigma=(p+q)\lambda_p-p\lambda-q\varpi$, where $\lambda$ is the true longitude of the asteroid, $\lambda_p$ the true longitude of the perturbing planet, $\omega$ the argument of perihelion, $\Omega$ the longitude of the ascending node, and $\varpi$ the longitude of perihelion. Libration or circulation of $\sigma$ over time was used to diagnose resonance capture and switching events}. 
As a result, 2010 JK$_{124}$ does not clearly fit into the standard dynamical categories and is therefore labeled as an "other TNO type". Objects belonging to different dynamical classes are expected to exhibit distinct levels of long-term orbital stability and sensitivity to planetary perturbations. Accordingly, the analysed TNOs have been grouped into three categories according to their long-term orbital stability: stable, unstable, and unresolved orbits. The results for each group are discussed in the following subsections.

\subsubsection{Stable orbits}
\label{Sect:TNO_Stable}
The two outer classical TNOs, 2010 KZ$_{39}$ and 2010 FX$_{86}$, are characterized by relatively low eccentricities and high orbital inclinations, having two of the three lowest eccentricities in our TNO sample (see Table~\ref{t7}--\ref{tab:starting-orbits}). With semi-major axes exceeding 45\,au and perihelia well beyond Neptune’s orbit, they are classical Kuiper Belt Objects, a subclass of TNOs. Our calculations indicate that their orbits are dynamically stable, with Lyapunov times, $T_L$, exceeding 1.1\,Myr. The long-term stability is further supported by the fact that, for all tested VAs, no convergence of $T_L(t)$ toward a constant value was observed within the integration timespan. The temporal evolution of $T_L(t)$ for five selected VAs, including the nominal orbit, is shown in the upper panel of Fig.~\ref{fig:evol_Lapunov_TNO}. 

In case of 2010 FX$_{86}$ the Lyapunov times is reported in the AstDyS-2 database, which lists $T_L \approx  2.5 \times 10^7$ years and it is consistent in our value. These findings align with previous studies showing that outer classical TNOs, particularly those on high-inclination orbits, remain dynamically stable over gigayear timescales \citep{LykawkaandMukai2005}.

In both cases, the $T_L$ distributions are symmetric around their median values and relatively narrow. For instance, the standard deviation of $T_L$ is  $\sim 5 \times 10^4$\,yr for 2010 KZ$_{39}$, and $\sim 10^5$\,yr for 2010 FX$_{86}$. The full distributions based on 1001~VAs are illustrated in Fig.~\ref{fig:Lapunov_TNO}, while statistical parameters describing these distributions are summarized in Table~\ref{tab:TNO-LT}.

\subsubsection{Unstable orbits}
\label{Sect:TNO_Unstable}
This group includes 2010 JK$_{124}$, (471143) Dziewanna, and 2010 JJ$_{124}$, originating from different dynamical regions as described in Sect.\ref{Sect:TNO_Intr}. Among all analysed TNOs, 2010 JJ$_{124}$ has the least well-determined orbit according to the $\Delta a / a$ criterion (column [10] in Tab.\ref{tab:starting-orbits}).

Despite their varied classifications -- from a scattered disc object in resonance (Dziewanna), to a Centaur (2010 JJ$_{124}$), and a non-resonant, unclassified TNO (2010 JK$_{124}$) -- all three exhibit long-term orbital instability in our numerical simulations. 
The first two, with perihelion distances of 35.7\,au and 32.5\,au, approach Neptune’s orbit closely enough to experience significant perturbations, while the third, with $q=23.6$\,au, crosses Neptune’s orbit, leading to even stronger interactions. In all cases, these encounters modify their orbital elements over the integration period. While instability is clearly visible for 2010 JJ$_{124}$ and 2010 JK$_{124}$, Dziewanna is less conclusive, as discussed below.

Table~\ref{tab:TNO-LT} shows that median $T_L$ estimates differ slightly depending on the applied method: $1.14$--$1.23 \times 10^5$\,yr for 2010 JK$_{124}$, $8.37$--$8.41 \times 10^4$\,yr for Dziewanna, and $8.48$--$8.59 \times 10^3$\,yr for 2010 JJ$_{124}$. Among these objects, only 2010 JK$_{124}$ has a Lyapunov time listed in the AstDyS-2 database, which reports $T_L \approx 1.2 \times 10^5$ yrs, in good agreement with our calculation. 

\begin{figure*}
	\centering
	\includegraphics[width=8.00cm]{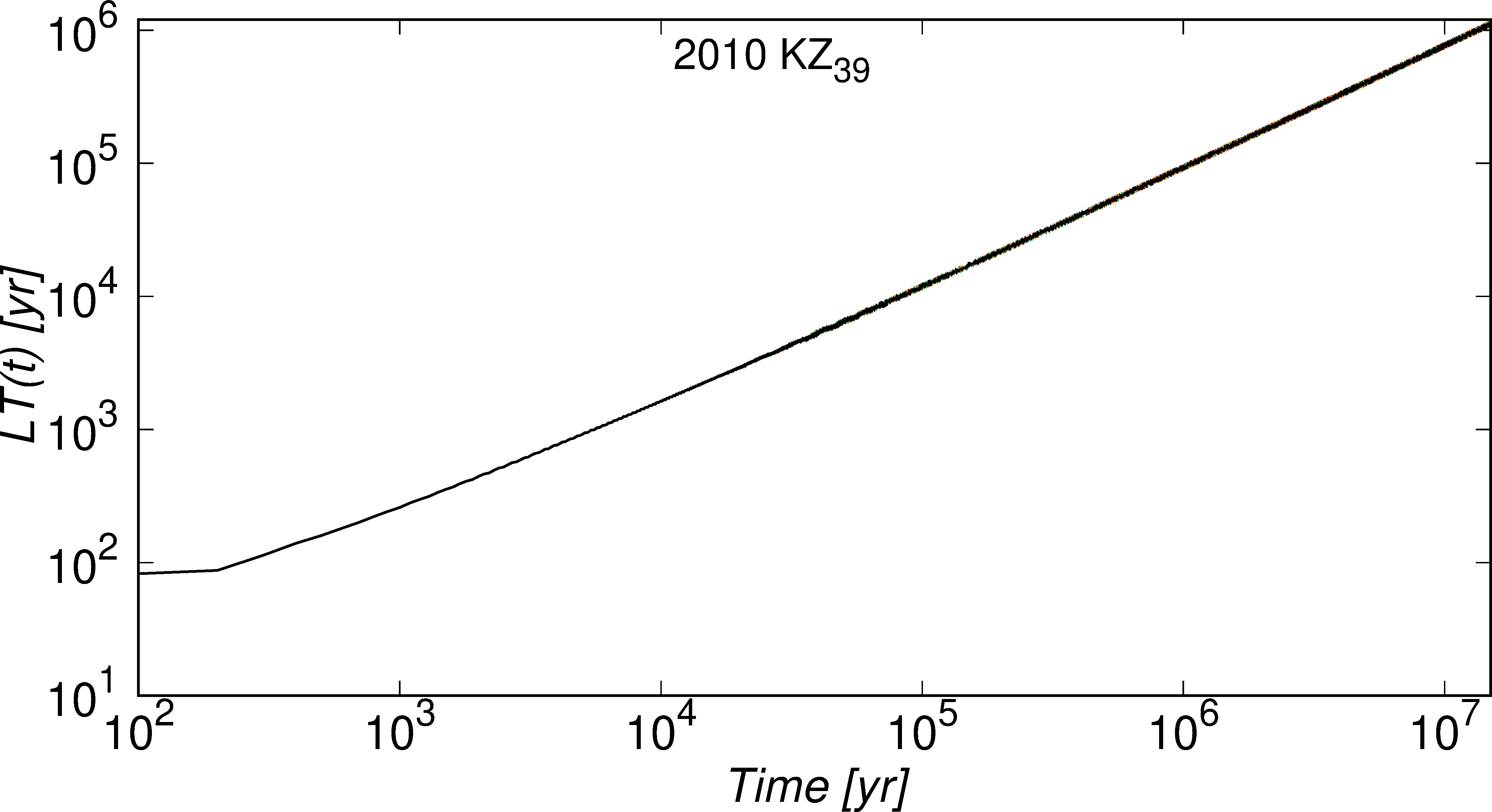}	
	\includegraphics[width=8.00cm]{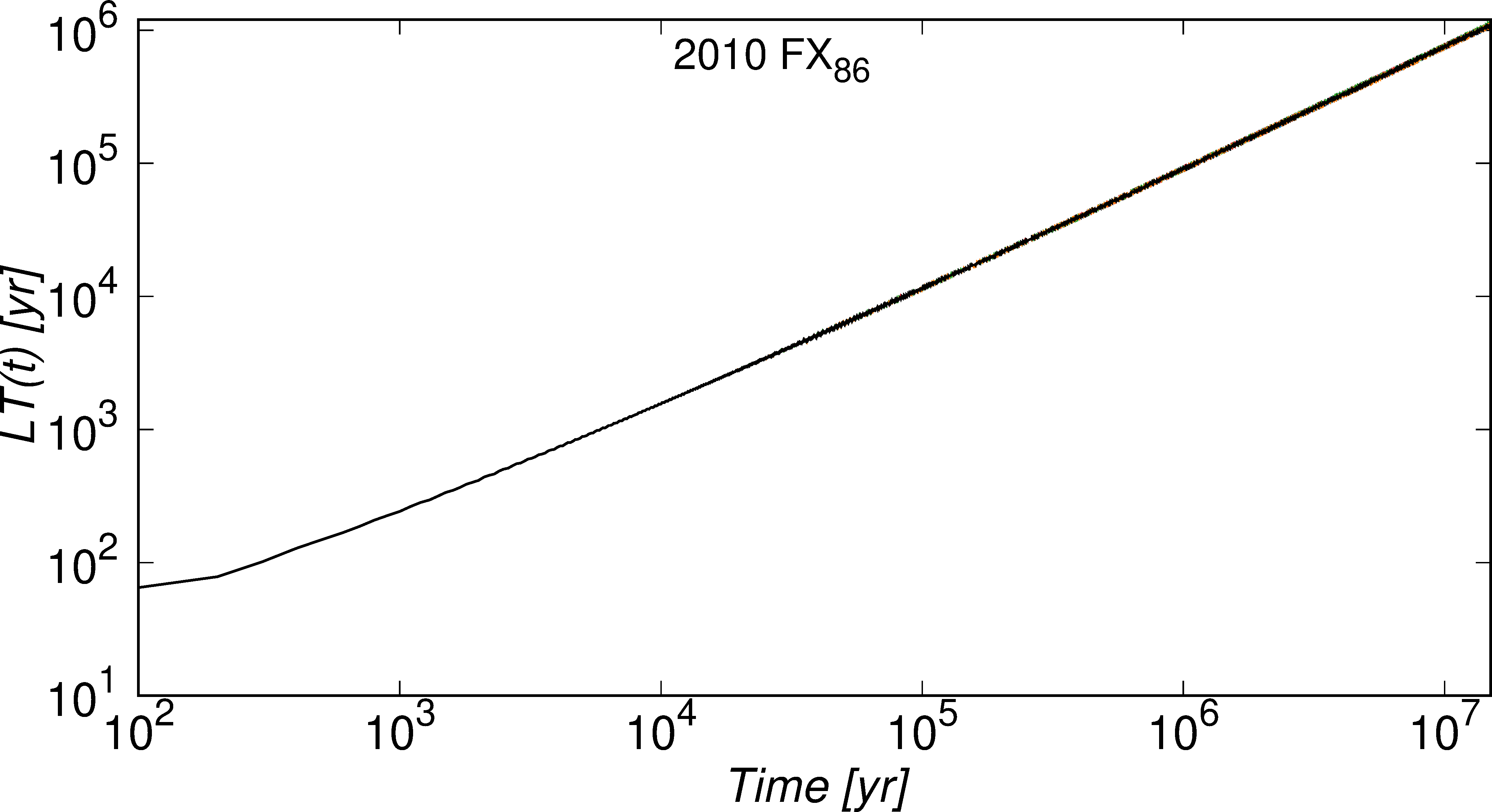}\\
	\includegraphics[width=8.00cm]{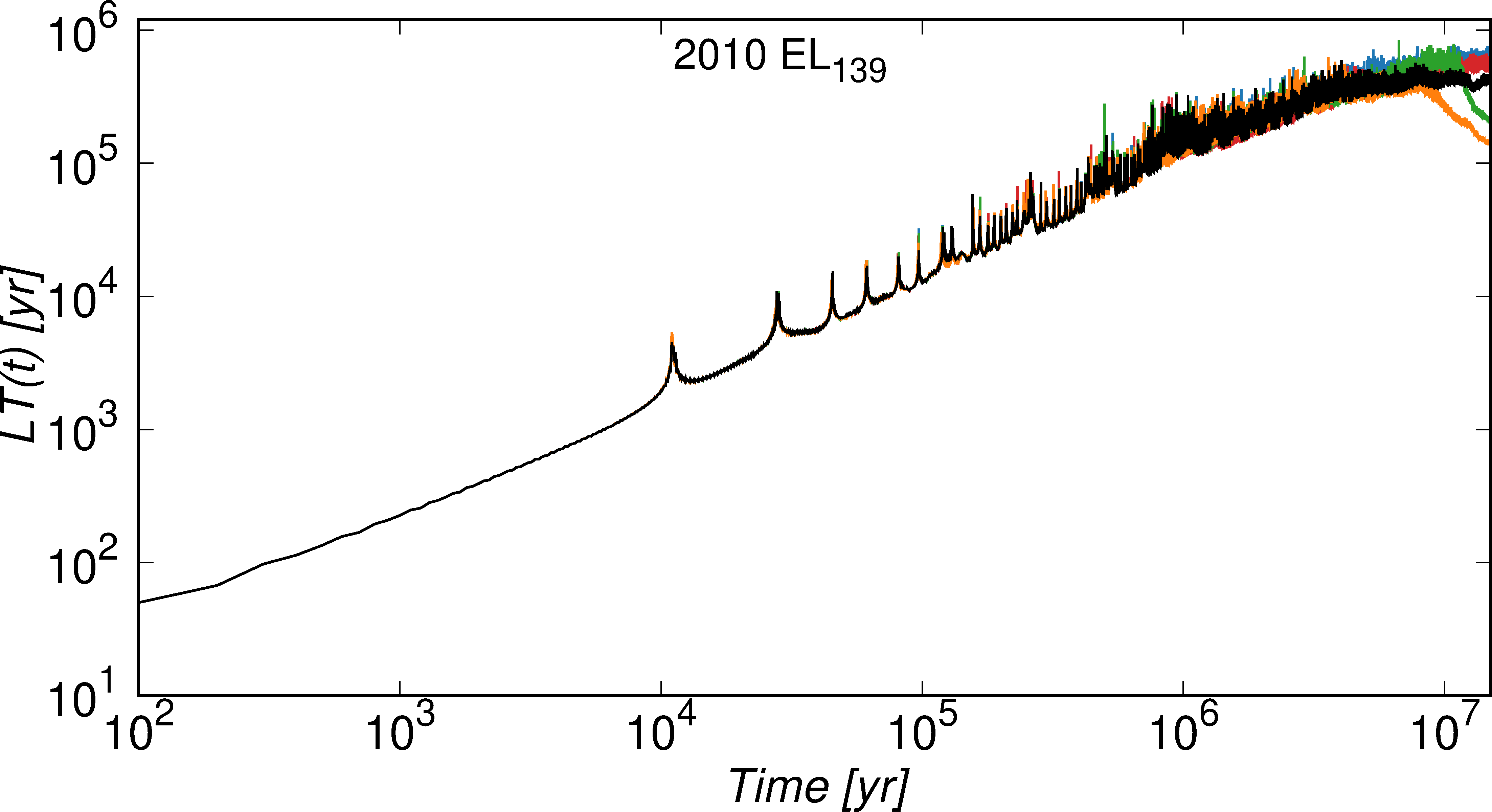}
	\includegraphics[width=8.00cm]{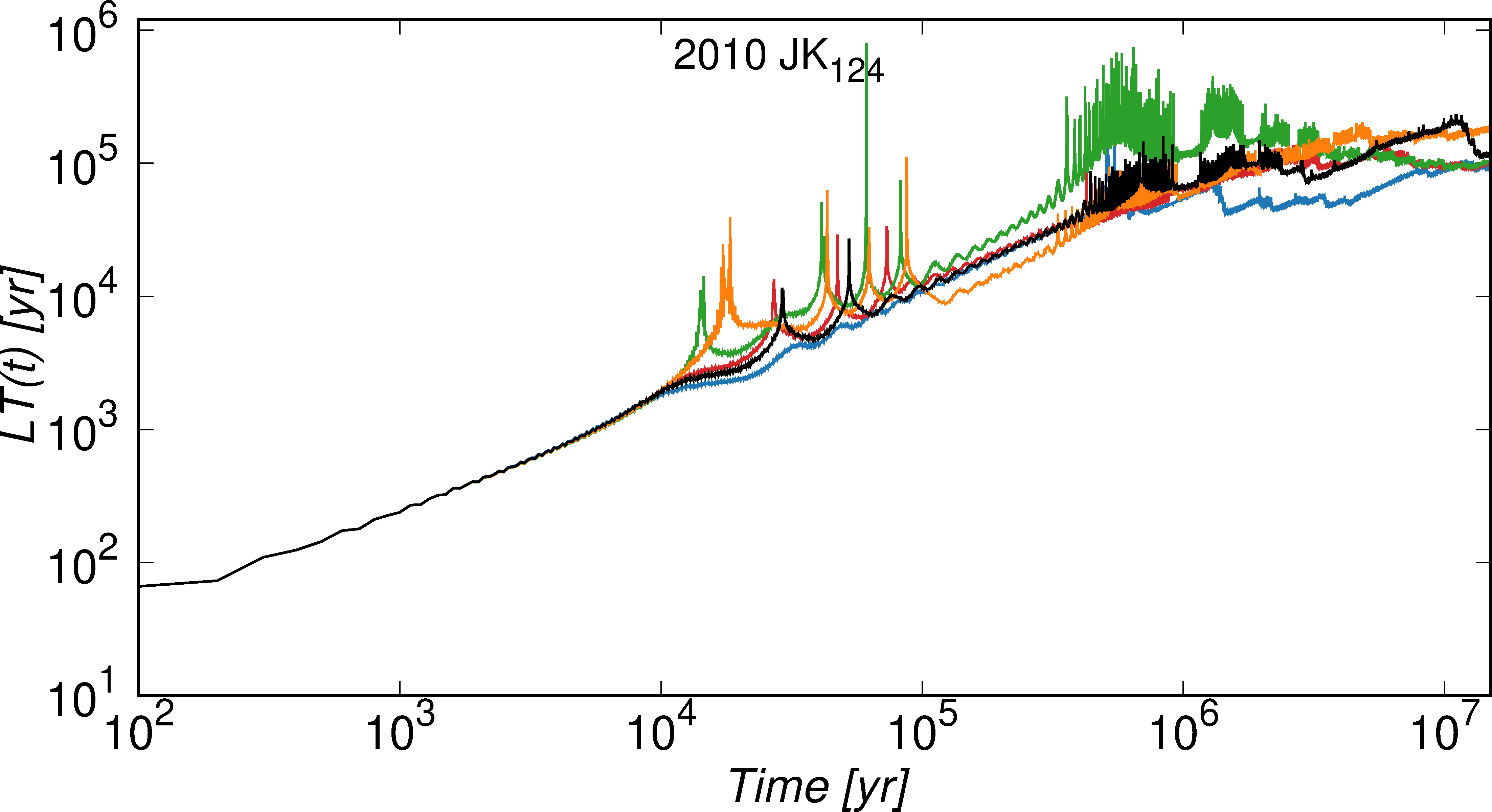}
	\includegraphics[width=8.00cm]{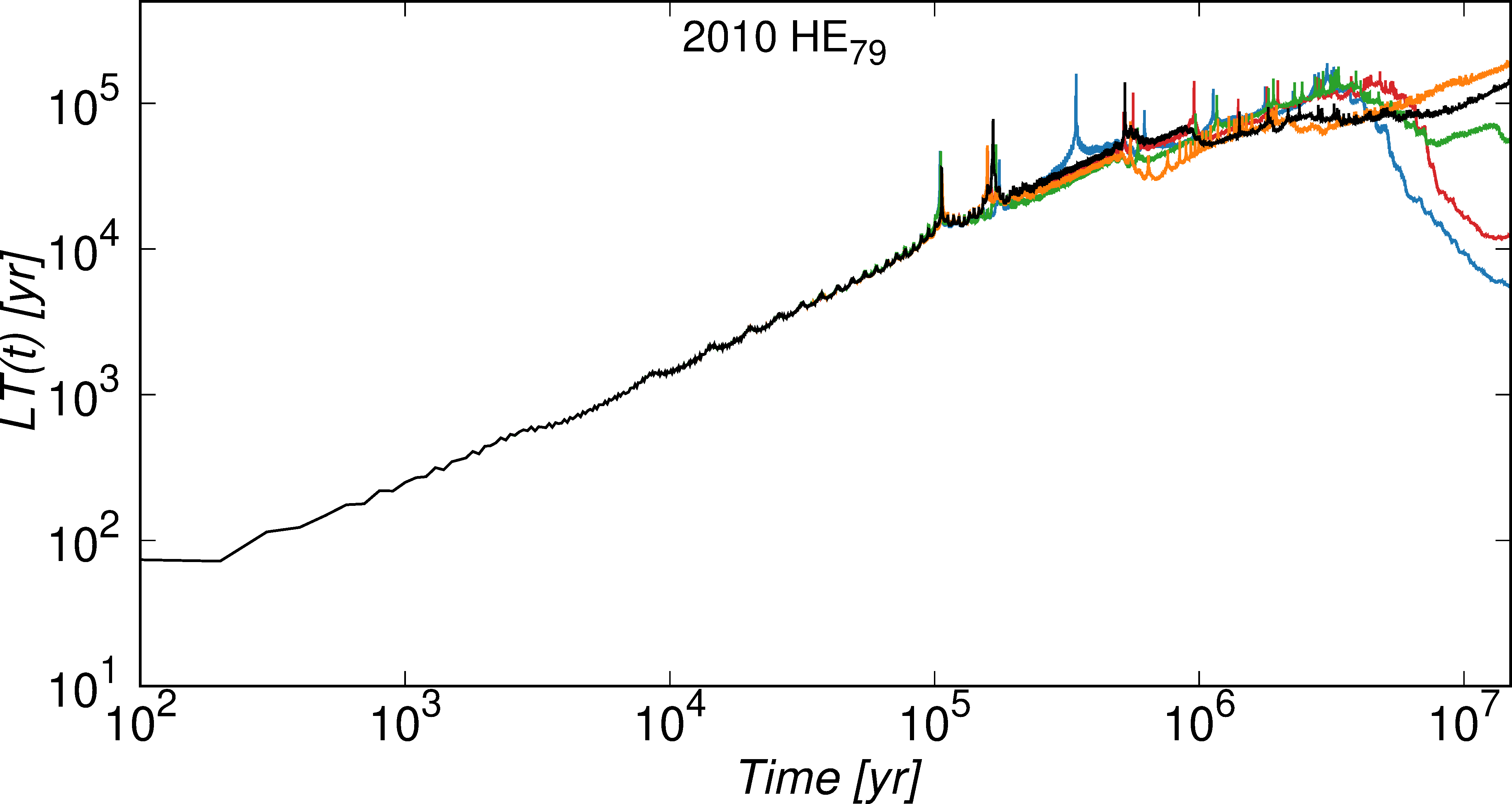}
	\includegraphics[width=8.00cm]{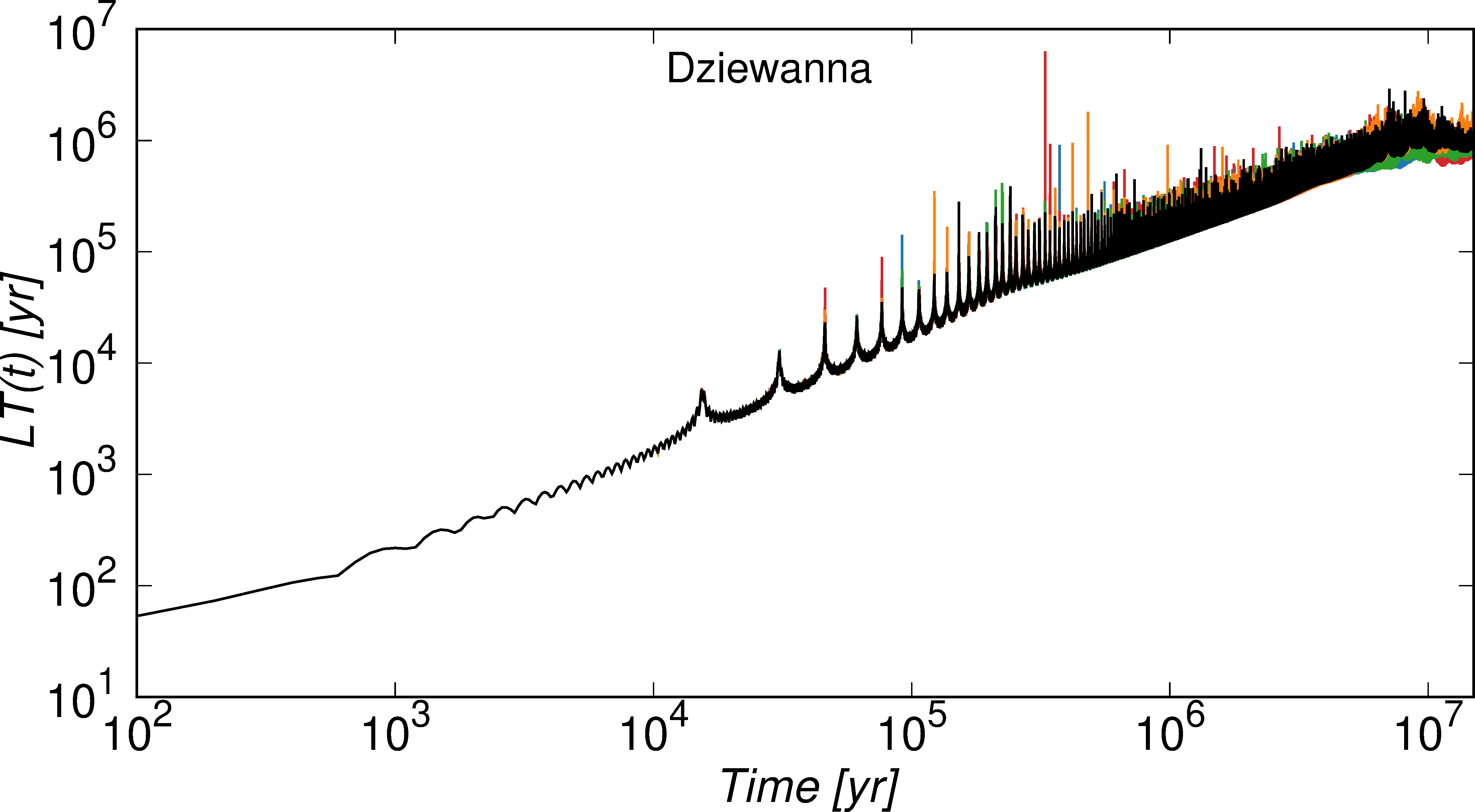}

   \hspace{8cm} \includegraphics[width=8.00cm]{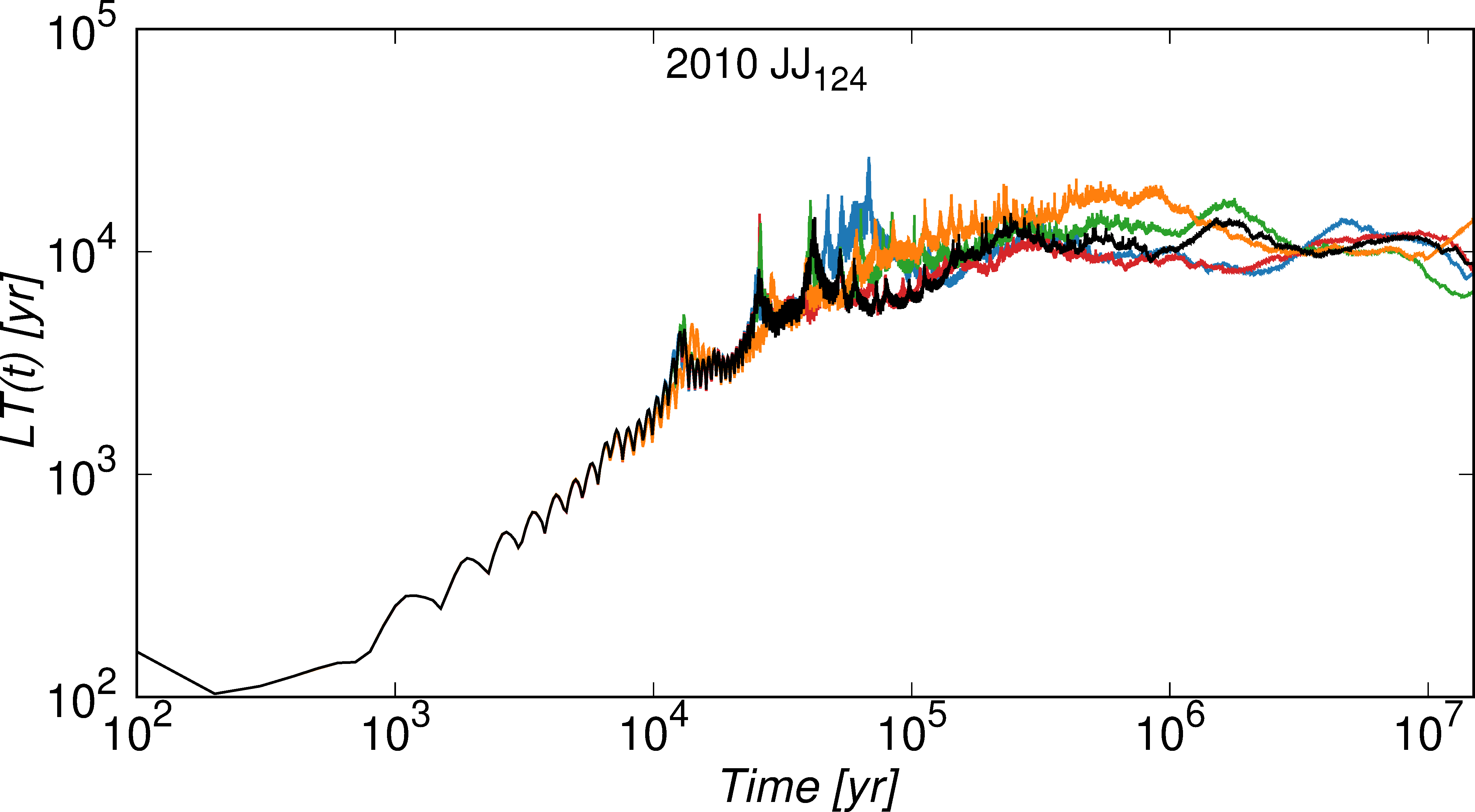}
	\caption{Time evolution of $T_L(t)$ for trans-Neptunian objects, showing the nominal orbit (black line) and four randomly selected virtual objects, calculated using the variational method.}
	\label{fig:evol_Lapunov_TNO}
\end{figure*}

As shown in Fig.~\ref{fig:evol_Lapunov_TNO}, the convergence of $T_L(t)$ toward a constant value is most evident for 2010 JJ$_{124}$, with the plot flattening after approximately $10^5$ years. For 2010 JK$_{124}$, convergence occurs later, around $10^6$ years, while Dziewanna only approaches convergence near the end of the integration period. Extending the integration time would therefore be necessary for a more complete assesment of $T_L(t)$ evolution and, ultimately, the final $T_L$ value. 
	
This ambiguity is consistent with \citet{Munoz-Gutierrezetal2025}, who investigated the dynamical evolution of Dziewanna (then designated provisionally as 2010 EK$_{139}$) as one of four large resonant and near-resonant TNOs. Their study, supported by our calculations, shows that Dziewanna resides near the 2:7 mean motion resonance with Neptune and may enter it  in the future, which plays a crucial role in shaping its secular evolution. Temporary libration of the argument of perihelion further indicates potential Kozai mechanism effects, coupling eccentricity and inclination over long timescales. This interplay of mean motion and secular resonances renders Dziewanna’s orbital dynamics complex: some trajectories  remain quasi-stable for extended periods, while others experience chaotic diffusion in semi-major axis and eccentricity. Consequently, the less conclusive nature of our Lyapunov time estimates for Dziewanna reflect the balance between temporary stability and instability. 

\begin{figure*}
	\centering
	\includegraphics[width=8.00cm]{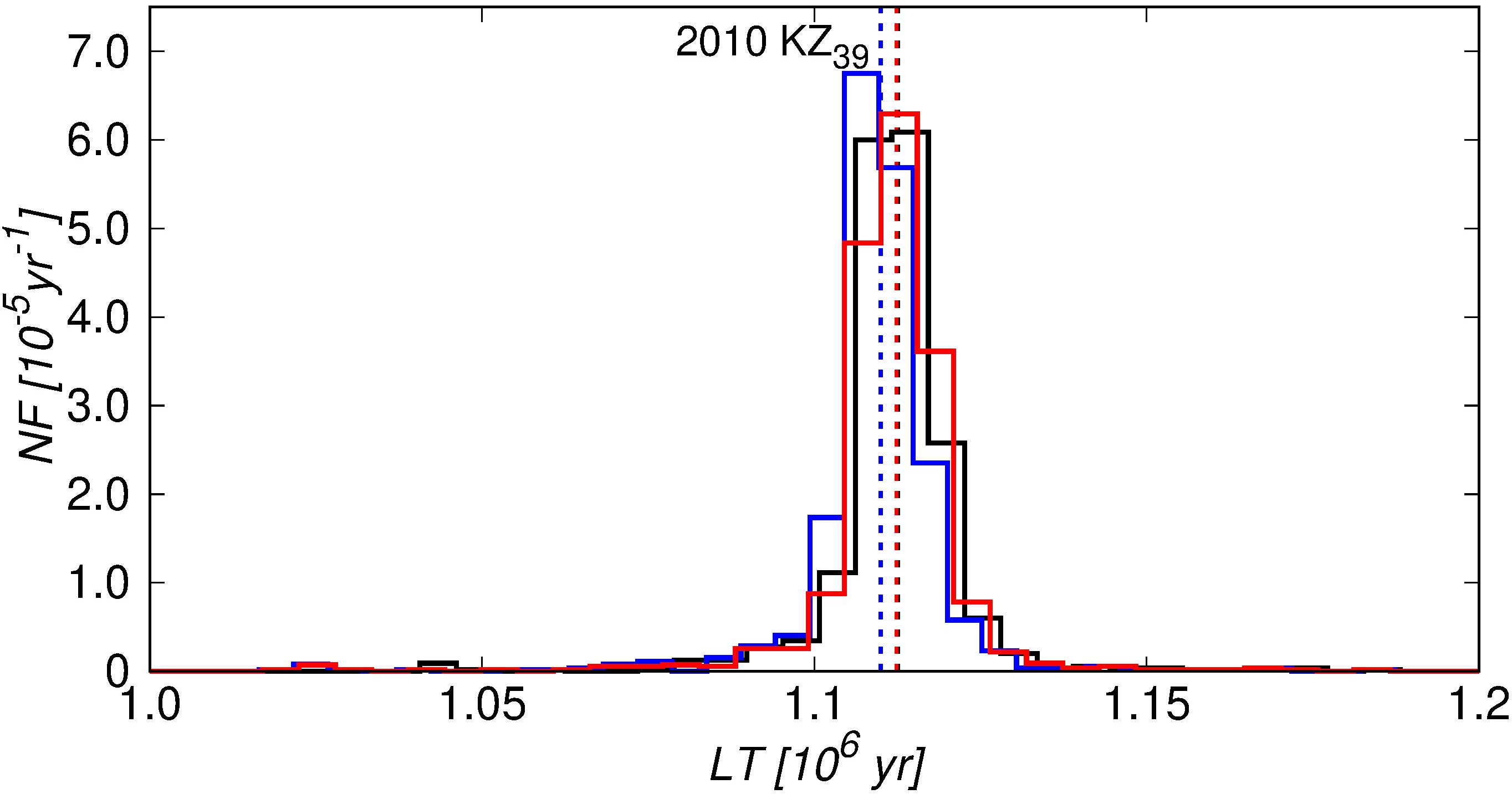}	
	\includegraphics[width=8.00cm]{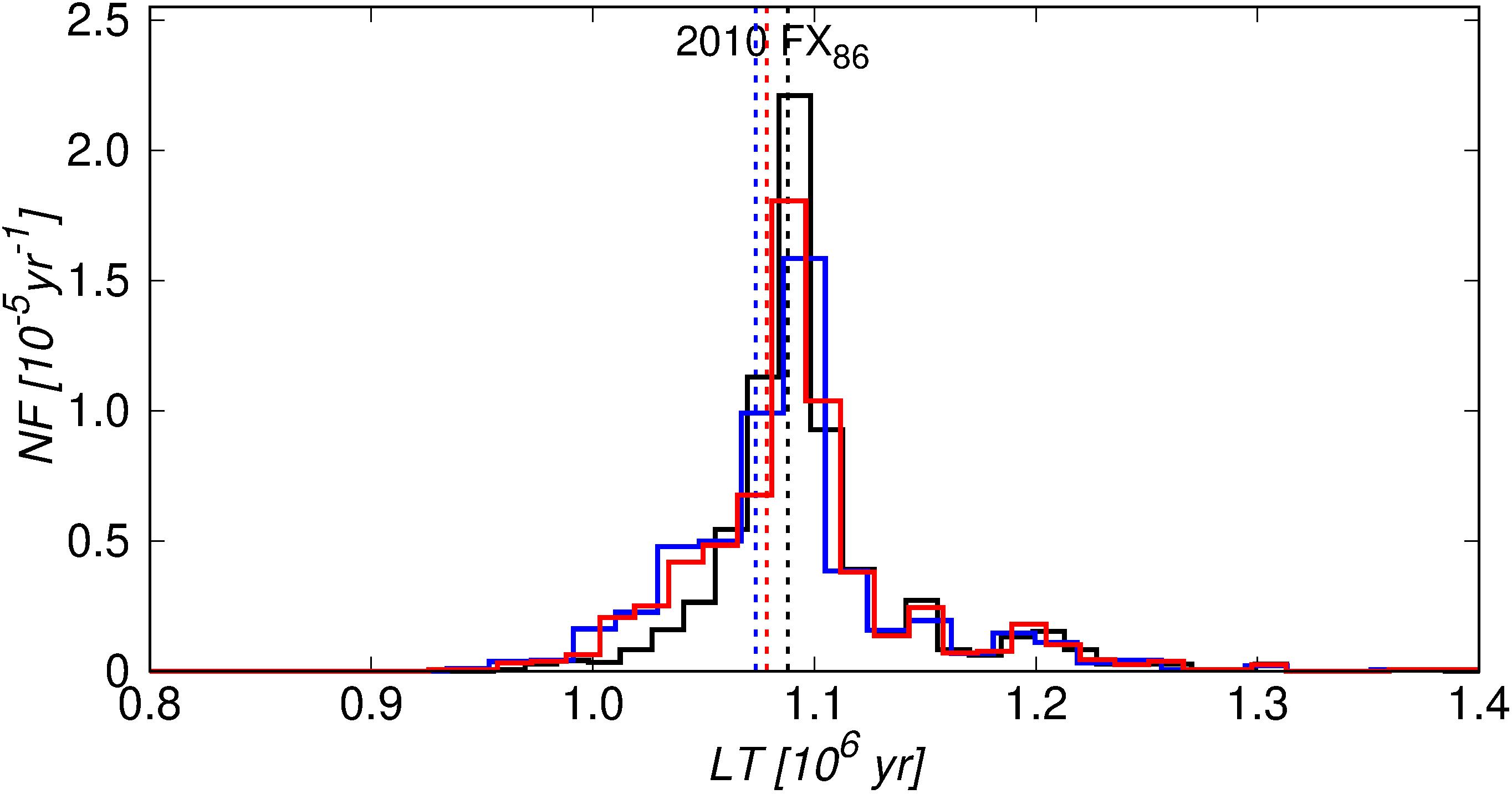}\\
	\includegraphics[width=8.00cm]{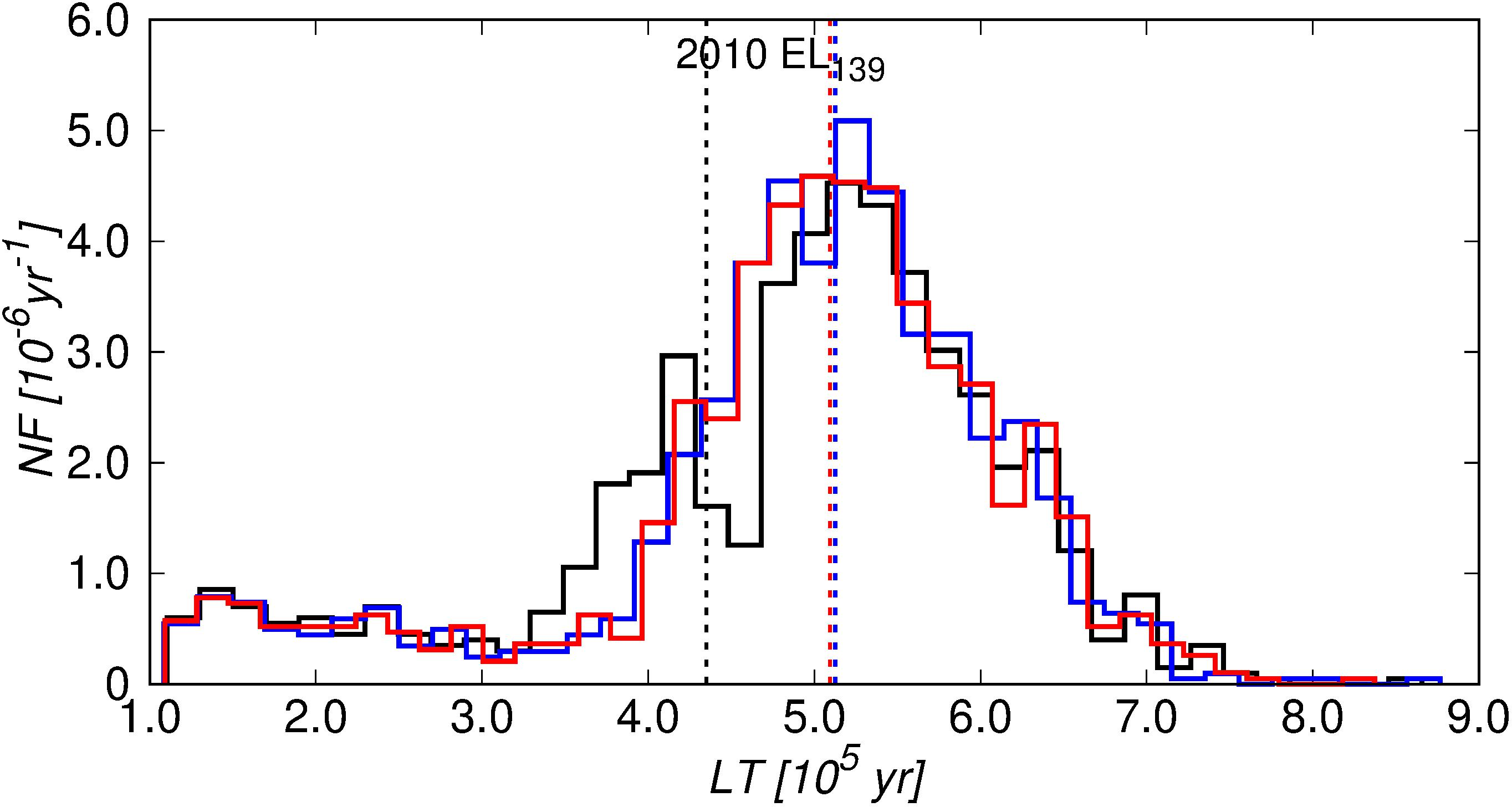}
	\includegraphics[width=8.00cm]{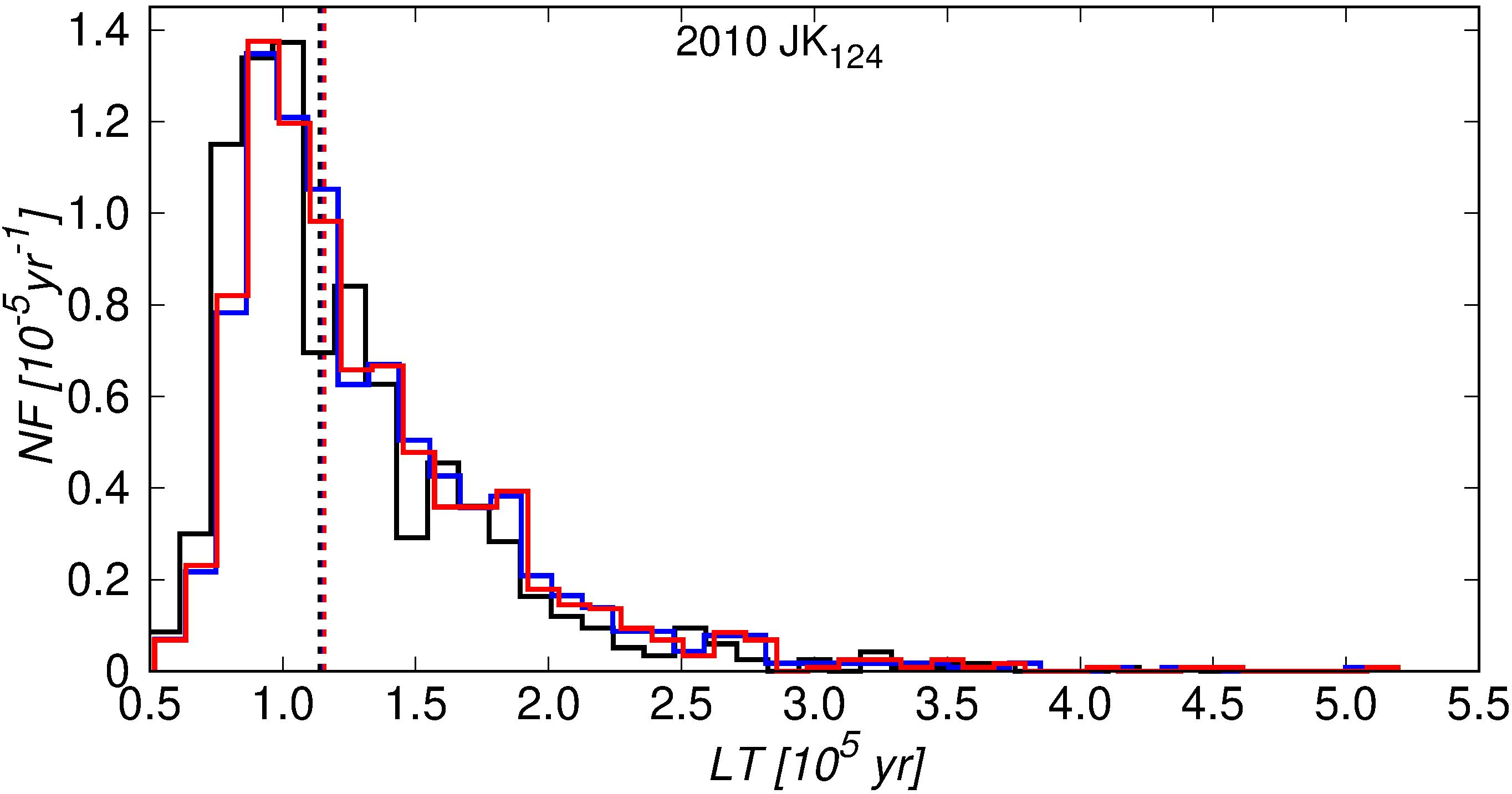}
	\includegraphics[width=8.00cm]{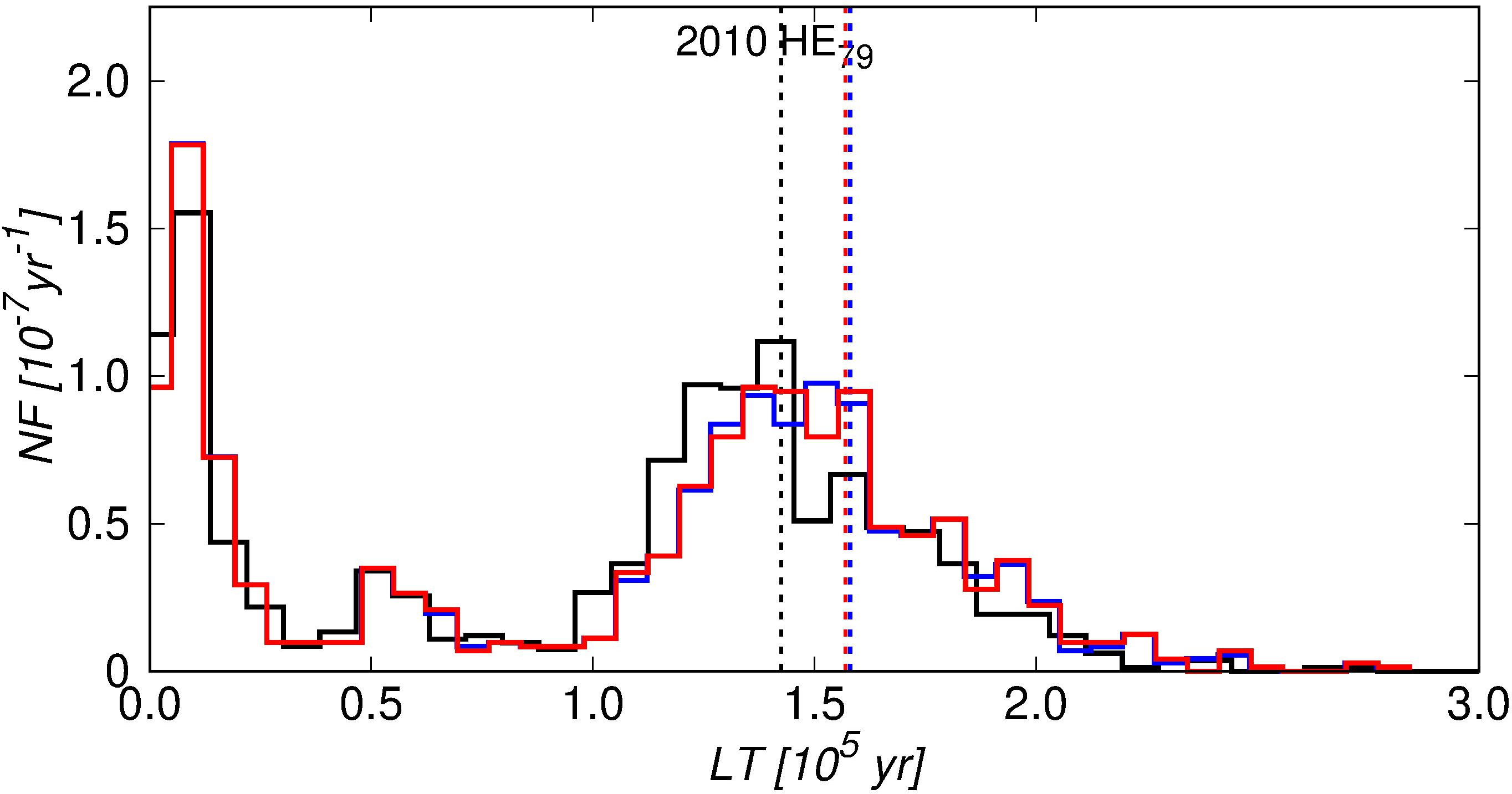}
	\includegraphics[width=8.00cm]{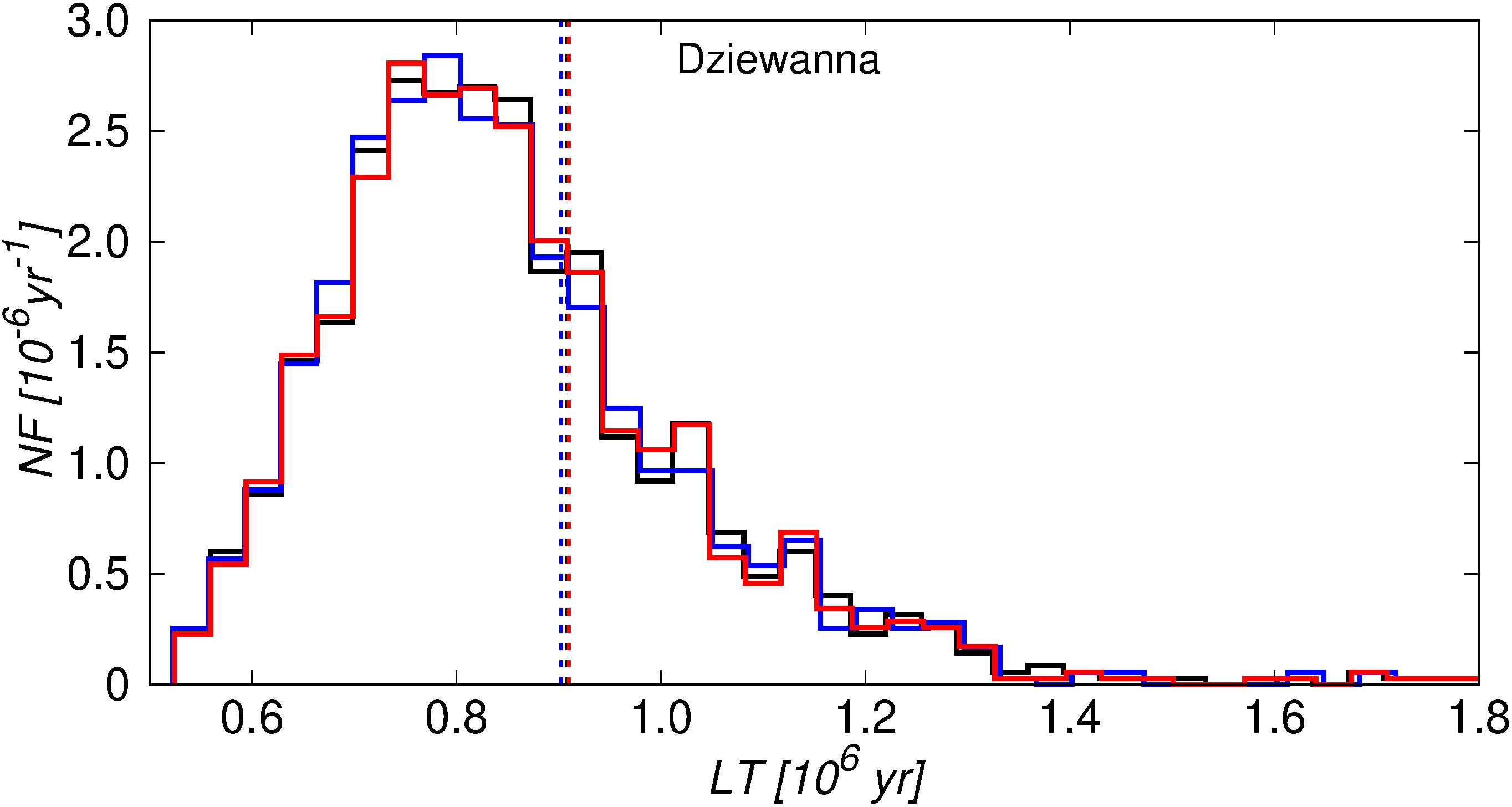}

\hspace{8cm}\includegraphics[width=8.00cm]{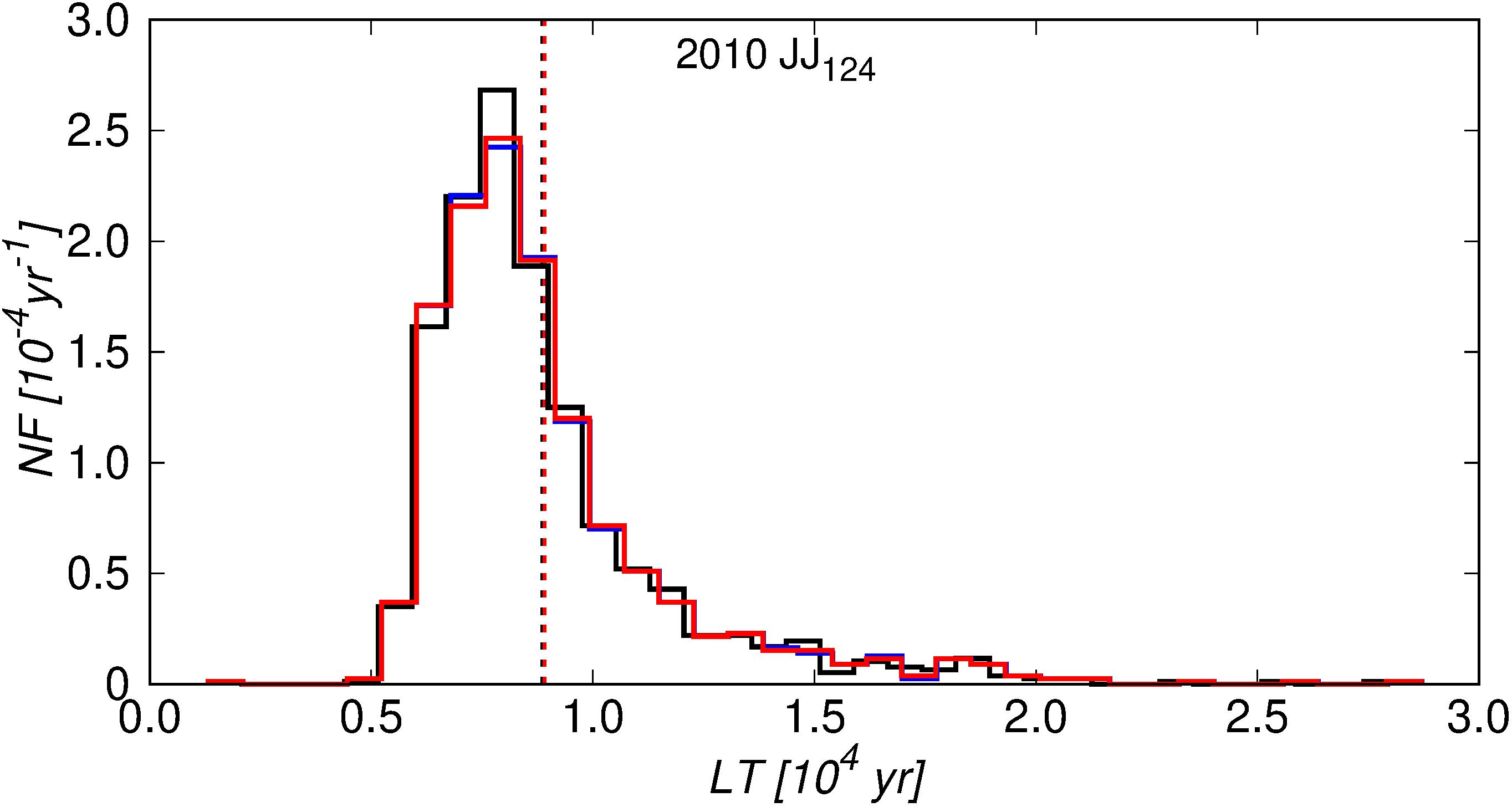}
	\caption{Statistical distribution of Lyapunov time calculated for the swarm of VAs after 15\,Myr for GR orbits representing TNO objects analysed in this study. The histograms show the calculation results performed by two methods: variational (black line) and neighbour trajectories method in two variants (blue and red lines). The $T_L$ for the nominal orbit is represented by the vertical dashed lines.}	\label{fig:Lapunov_TNO}
\end{figure*}

Figure~\ref{fig:Lapunov_TNO} illustrates that  $T_L$ distributions for the VAs are right-skewed for all three objects, with the corresponding skewness values listed in Table~\ref{tab:TNO-LT}. Skewness is particularly pronounced for 2010 JK$_{124}$, where some VAs exhibit $T_L$ values up to five times larger than the median being slightly above $10^5$ years. Dziewanna and  2010 JJ$_{124}$ exhibit narrower, more concentrated VA distributions. 

\subsubsection{Unresolved orbits}
\label{Sect:TNO_Unresolved}
This section discusses 2010 EL$_{139}$ and 2010 HE$_{79}$, which, according to the $\Delta a / a$ criterion (column [10] in Tab.~\ref{tab:starting-orbits}), belong to the best-determined orbits in our sample. Nevertheless, their long-term orbital stability remains uncertain, as some VAs exhibit complex dynamical behaviour, showing both stability and potential transitions to less stable regimes. This ambiguity makes a clear classification of their orbital character challenging. 

The orbit of 2010 EL$_{139}$ appears stable for most of VAs. The estimated median of $T_L$ exceeds 0.5 Myr across all three methods and $IQR$ ranges from about $1.2\times 10^5$ to $1.5\times 10^5$, depending on the method. The AstDyS-2 database lists $T_L \approx 5.3 \times 10^5$\,yr for this object, which is similar to our results. As shown in Fig.\ref{fig:Lapunov_TNO}, $T_L$ values between $\sim 0.1$ and $\sim 0.3$\,Myr form a nearly flat distribution, with a slight upward trend toward shorter timescales. This part of the histogram consists of only about 10\% of all VAs. The values of $T_L(t)$ for this fraction increase over time, reaching up to about 10 Myr, after which a decrease in $T_L$ can be observed. This behaviour can be explained by the fact that these objects may transition into a different dynamical regime (as discussed in the following paragraph), possibly characterized by greater dynamical instability, unlike other VAs, for which $T_L$ continues to increase over time. Representative examples of such divergent $T_L(t)$ behaviours are illustrated in Fig.~\ref{fig:evol_Lapunov_TNO}.

Further insight into the dynamical state of 2010 EL$_{139}$ is provided by its resonant behavior. We have discovered that this object is in a 2:3 mean-motion resonance with Neptune (i.e., the main resonance angle associated with this resonance oscillates with a large amplitude of $\sim$150\degr). However, for VAs with relatively small $T_L$, transitions from libration to circulation of the resonant angle do occur, typically after integration times longer than 5\,Myr. The frequency of such transitions decreases with increasing Lyapunov time, and for VAs with $T_L \gtrsim 3 \times 10^4$\,yr, no cases of circulation were observed. For this particular object, the observed correlation between $T_L$ and resonant behaviour suggests that deeper trapping within the 2:3 mean-motion resonance is associated with larger Lyapunov times, and thus with increased dynamical stability in the Lyapunov sense.

Similar to 2010 EL$_{139}$, the orbital behaviour of 2010 HE$_{79}$ also shows signs of partial stability combined with significant dynamical diversity among virtual clones. In this case, we found that the median $T_L$ is approximately $1.27$–$1.33 \times 10^5$\,yr depending on the method used to estimate this parameter (see Tab.\ref{tab:TNO-LT}). As shown in Fig.\ref{fig:Lapunov_TNO}, the dispersion of $T_L$ values for this asteroid is quite large, with the interquartile range ($IQR$) even exceeding the median value. From this figure, we can also observe that the $T_L$ values of VAs cluster around three distinct maxima: approximately 4 kyr ($\sim$29\% of VAs), 50\,kyr ($\sim$9\%), and 0.13 Myr ($\sim$62\%). VAs belonging to the first two groups likely exhibit chaotic behaviour, whereas for those in the third group we found no convergence of $T_L(t)$ over time. This suggests that their orbits are either stable or that their true $T_L$ values are significantly larger than the current integration window allows us to resolve. A representative example of such divergent behaviours in the evolution of $T_L(t)$ is illustrated in Fig.~\ref{fig:evol_Lapunov_TNO}. 
	
We did not examine the dynamical evolution of all 1,001~VAs for this object, but we randomly selected a dozen of them and traced their evolution. Our analysis shows that the orbits of VAs clustered around each of these maximum $T_L$ values evolve in different ways. The first group, clustered around $T_L \sim 4$ kyr, exhibits irregular orbital evolution: after some time the semi-major axis becomes erratic, and, more importantly, the perihelion distance $q$ decreases below Neptune’s semi-major axis to values of order $\sim$25\,au. This causes close encounters with Neptune, resulting in strong perturbations to their orbits and a decrease in stability. The second group, with $T_L \sim 5 \times 10^4$ years, represents transitional orbits: here, $a$ usually remains regular for most of the integration time, with sudden changes occurring only at the end. In this case, $q$ typically approaches Neptune's orbit but does not fall below it, resulting in slightly greater stability compared to the first group. Finally, the third maximum, at $T_L \sim 1.3 \times 10^5$\,yr, is associated with the most regular behaviour: $a$ remains regular, $q$ stays above Neptune’s semi-major axis (often even increasing), and the orbits in this group interact least with Neptune, making them the most stable of the three. 

Compared with 2010 EL$_{139}$, where the signs of instability are less directly linked to specific perihelion behaviour, the case of 2010 HE$_{79}$ shows a clearer dependence on Neptune’s perturbations. This highlights the particularly strong role of Neptune in shaping its long-term dynamical evolution. Finally, we would like to stress that these two objects exhibit a left-skewed, asymmetric distribution of $T_L$, which distinguishes them from the objects in the other two groups\footnote{The only exception is 2010 FX$_{86}$, analyzed in Sect.~\ref{Sect:TNO_Stable}, for which we also observed a left-skewed distribution of $T_L$, but this occurred only when using the renormalization methods.}. This can be explained as follows: most VAs have relatively large $T_L$ and therefore more stable orbits. On the other hand, the tail of the distribution extends towards smaller $T_L$ values, meaning that there is a smaller but significant proportion of clones that evolve in less stable dynamical regimes, as discussed above.

\begin{table*}
\caption{\label{tab:TNO-LT}
Statistical parameters describing the distribution of Lyapunov times calculated for 1001 VAs of the trans-Neptunian objects (TNOs) analysed in this study. The listed parameters include the first, second (median), and third quartiles ($Q_1$, $Med$, $Q_3$), the interquartile range ($IQR$), skewness ($A_Q$), the Lyapunov time of the nominal orbit ($T_L^\mathrm{nom}$), and the relative spreads of the median ($\delta_{Med}$) and nominal ($\delta_{nom}$) values. In column 2, symbol 'V' refers to the Lyapunov time estimated using the variational method, 'N1' and 'N2' refer to two variants of the renormalization method. All dimensional quantities in the table are expressed in Julian years. See the main text for further details.}
\centering
\begin{tabular}{ccccccccc}
\hline \hline 			 
\multicolumn{2}{c}{\backslashbox{Par/method}{Object}} & 2010 KZ$_{39}$   & 2010 FX$_{86}$  & 2010 EL$_{139}$  & 2010 JK$_{124}$ & (471165) 2010 HE$_{79}$ & (471143) Dziewanna & 2010 JJ$_{124}$ \\ 
\hline  \hline 
 & V&1.1149$\times 10^6$   &1.0991$\times 10^6$  &5.1890$\times 10^5$  &1.1408$\times 10^5$  &1.2671$\times 10^5$  &8.4073$\times 10^5$  & 8.4805$\times 10^3$  \\ 
$Med$ & N1&1.1122$\times 10^6$  &1.0971$\times 10^6$ &5.2131$\times 10^5$  &1.2244$\times 10^5$  &1.3344$\times 10^5$ &8.3712$\times 10^5$ &8.5958$\times 10^3$ \\ 
 & N2&1.1146$\times 10^6$  &  1.0989$\times 10^6$&5.1901$\times 10^5$  &1.2280$\times 10^5$  &1.3317$\times 10^5$ &8.4149$\times 10^5$ & 8.5947$\times 10^3$\\ 
 \hline
  & V&7.700$\times 10^3$    & 2.7300$\times 10^4$ & 1.5262$\times 10^5$ & 5.3489$\times 10^4$ &1.2785$\times 10^5$ &2.0508$\times 10^5$ &2.454$\times 10^3$ \\ 
IQR & N1& 7.630$\times 10^3$ &3.9710$\times 10^4$  & 1.2082$\times 10^5$ &5.9844$\times 10^4$  & 1.3748$\times 10^5$&2.0537$\times 10^5$ &2.524$\times 10^3$ \\ 
 & N2&7.640$\times 10^3$  &3.7260$\times 10^3$  &1.2183$\times 10^5$  &5.9087$\times 10^4$  &1.3744$\times 10^5$  &2.0509$\times 10^5$&2.533$\times 10^3$\\ 
  \hline
 & V&0.103896   &0.152381  &-0.218963  &0.342377   &-0.626074   & 0.118664  &0.21194   \\ 
$A_Q$ & N1&0.0930537   &-0.231932   &-0.0452409   &0.313983   &-0.606956   &0.123996   &0.197372   \\ 
 & N2&0.0968586   &-0.183038   &-0.0257324   &0.303468   &-0.603346   &0.123265   &0.204414   \\ 
  \hline
   & V&1.1126$\times 10^6$ &1.0879$\times 10^6$ &4.3491$\times 10^5$  &1.1367$\times 10^5$  &1.4246$\times 10^5$ & 9.0788$\times 10^5$&8.8559$\times 10^3$
\\ 
$T_L^\mathrm{nom}$ & N1&1.1099$\times 10^6$  &1.0735$\times 10^6$  &5.1250$\times 10^5$  &1.1521$\times 10^5$&1.5804$\times 10^5$ &9.0218$\times 10^5$ &8.9113$\times 10^3$
 \\ 
 & N2&1.1123$\times 10^6$ &1.0784$\times 10^6$  &5.0921$\times 10^5$ &1.1564$\times 10^5$ &1.5700$\times 10^5$ &9.0986$\times 10^5$ &8.9114$\times 10^3$
  \\
\hline 
$\delta_{Med}$ &   &2.4506$\times 10^{-3}$  &1.8665$\times 10^{-3}$ &4.4272$\times 10^{-3}$& 3.0390$\times 10^{-3}$ &2.0595$\times 10^{-3}$&5.2097$\times 10^{-3}$&1.1686$\times 10^{-4}$
 \\
\hline 
$\delta_{nom}$ & & 2.4046$\times 10^{-3}$ & 1.3373$\times 10^{-2}$ &6.7873$\times 10^{-3}$ &3.7234$\times 10^{-3}$ &6.8181$\times 10^{-3}$ &8.4759$\times 10^{-3}$ &2.9237$\times 10^{-6}$
\\ 
\hline \hline 
$T_L^\mathrm{lit}$ &AstDyS-2 & - & 2.50$\times 10^7$ & 5.3476$\times 10^5$  &1.1848$\times 10^5$
  & -- &  -- & -- 
 \\ 
\hline 
\end{tabular} 
\label{tab: Lapunov_TNO}
\end{table*}

\subsection{Outer MBA} 
\label{Sect:oMBA}
Outer main-belt asteroids (MBAs) are typically defined as those with semi-major axes larger than 2.82 au. This region lies beyond the 5:2 mean motion resonance (MMR) with Jupiter (at $\sim$2.82 au) and extends beyond the 2:1 MMR resonance with that planet (at $\sim$3.28 au). Dynamically, the outer main belt is characterized by a relatively high degree of orbital excitation, including elevated eccentricities and inclinations compared to the inner and middle belts. This structure results from the long-term gravitational perturbations caused primarily by Jupiter and Saturn, as well as the influence of mean motion and secular resonances. The locations of the main MMRs with Jupiter are indicated in Fig.~\ref{fig:MBA_classes}.

Among the five outer Main Belt asteroids analysed in this work, (522) Helga is located in the Cybele region, beyond the 2:1 MMR with Jupiter, and has been associated with a local collisional group often referred to as the Helga group \citep{Carrubaetal2015}. (1328) Devota also resides in the Cybele region, although its membership in a specific collisional family is less firmly established. The remaining objects — (2311) El Leoncito, (3095) Omarkhayyam, and (1144) Oda — are not commonly associated with any major collisional families.

\subsubsection{Overview of Orbital Characteristics of the Analysed oMBAs}\label{Sect:oMBA_Intr}
\begin{figure*}
	\centering
	\includegraphics[width=18.00cm]{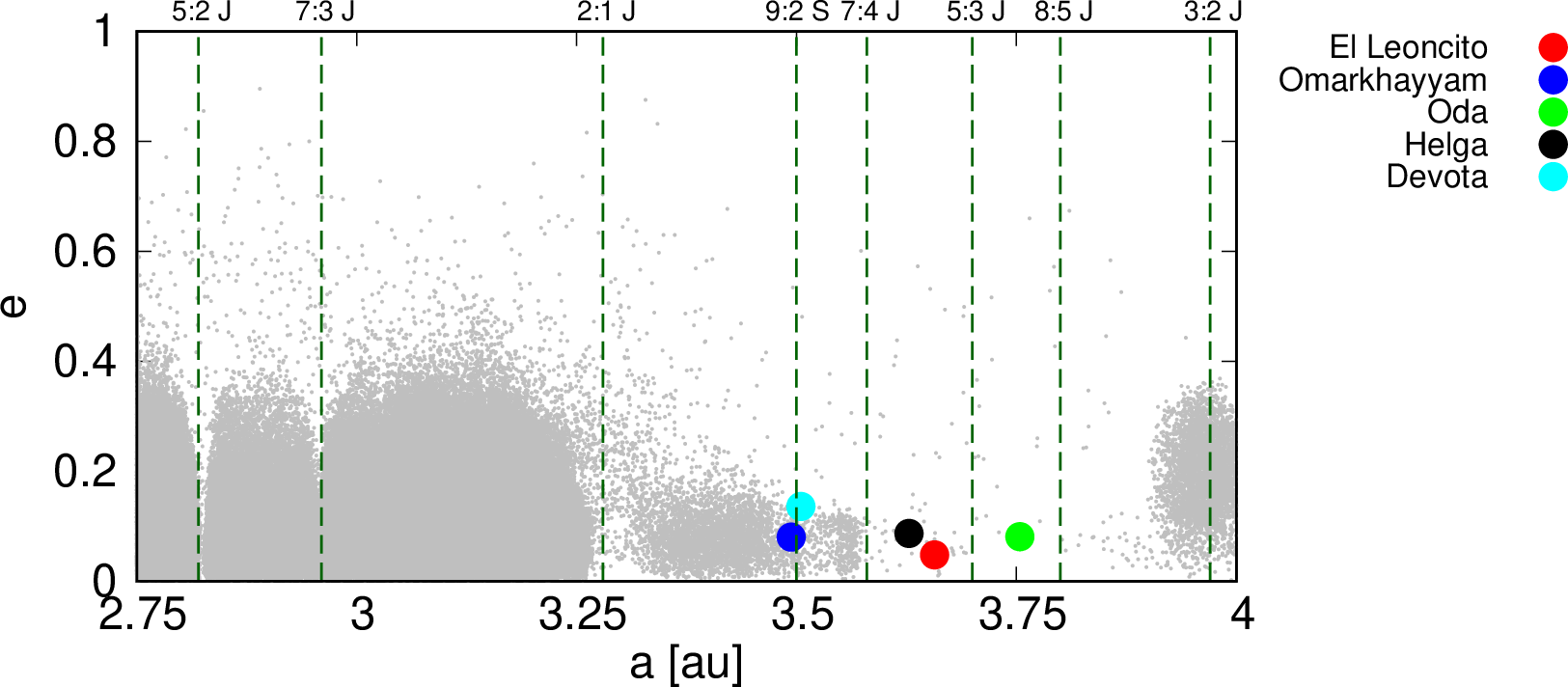}
	\caption{Distribution of the analysed Main Belt asteroids in the $a$–$e$ plane, with the main mean motion resonances. These resonances are represented on the plot by a vertical dashed green lines. Grey points illustrate the overall distribution of numbered Main Belt asteroids only, based on data retrieved from the AstDys-2 webpage on 14 January 2026.}
		\label{fig:MBA_classes}
\end{figure*}

In this section, we investigate the Lyapunov time and stability of five asteroids from the outer Main Belt: El Leoncito, Omarkhayyam, Oda, Helga, and Devota. Although all belong to the same dynamical region, the orbital parameters of these bodies — including semi-major axes, eccentricities, and inclinations — place them in different parts of the outer belt, offering a broad view of its dynamical diversity. In particular, their semi-major axes lie between 3.50 and 3.75 au, with inclinations ranging from low to moderate values (3–10 degrees) and eccentricities spanning 0.04–0.14, which affects their long-term orbital behaviour. The distribution of the analysed Main Belt asteroids in the $a$–$e$ plane, along with the main mean motion resonances indicated by vertical dashed green lines, is shown in Figure \ref{fig:MBA_classes}. All these asteroids have been observed in more than 30 oppositions, ensuring that their orbits are generally very well determined. Attempts to derive NG orbits were also made, but in each case the results were inconclusive.

The Lyapunov time of objects in this group was previously estimated by \citet{Murisonetal1994}, \citet{HolmanandMurray1996} and \citet{NovakovicandRadovic2019}; however, they used different methods and distinct models of the Solar System. \citet{Murisonetal1994} used the elliptic restricted three-body (ERTBP) problem, with Jupiter as the secondary mass. They performed integrations up to $10^5$ Jupiter years ($\sim$1.2\,Myr), and the Lyapunov time was estimated using the renormalisation method. \citet{HolmanandMurray1996} also performed the estimation in the ERTBP model using two different approaches: the variational method and the renormalization method. In this case, the authors stopped the simulations when they reached $1.6 \times 10^6$  Jupiter periods (19\,Myr). \citet{NovakovicandRadovic2019} estimated the Lyapunov time from the integration up to 10\,Myr. They used the Solar System model with seven planets, excluding Mercury. Values of the $T_L$ is also reported in the Asteroid Dynamic Site (AstDyS-2) webpage. In this case, the Lyapunov time was estimated after performing forward integrations for a period of 2\,Myr. Only in the case of the Helga asteroid, the calculation time was 10\,Myr. Based on our $T_L$ estimates for these five MBA, we are able to classify them into two dynamical categories: objects with stable orbits and those with unstable orbits. In the following sections, we present a detailed analysis of the Lyapunov times for each of them.

\subsubsection{Stable orbits}
\label{Sect:oMBA_Stable}
Stable orbits were found for El Leoncito and Omarkhayyam, as indicated by the analysis of the time evolution of the $T_L(t)$ parameter. For both asteroids, this parameter does not converge to a constant value, as shown in Fig.~\ref{fig:evol_Lapunov_oMBA}.

As seen in Table~\ref{tab:MBA-LT}, for El~Leoncito the lower limit of the median of $T_L$ estimate, determined by all three methods, is approximately 1.2\,Myr. A slight difference is observed only in the case of the N1 method compared to the other two. The median $T_L$ of all VAs is largely consistent with the $T_L$ value for the nominal orbit. The histograms of $T_L$ distributions are slightly right-skewed, as also shown in Fig.~\ref{fig:Lapunov_oMBA}. It is worth noting that, for all three methods, the interquartile range ($IQR$) of the obtained $T_L$ values is small, on the order of 20,000\,yr, reflecting a low dispersion of $T_L$ estimates for individual VAs. Our calculations are in good agreement with the $T_L$ estimates presented in the AFP and AstDys-2 databases. By contrast, the results obtained by \citet{Murisonetal1994} and \citet{HolmanandMurray1996} suggest that the orbit of this object is unstable, with $T_L$ estimates on the order of 5,000--6,000\,yr. However, it should be empasized that their estimates were derived using a simplified model of the Solar System, in which Jupiter was the only planet included.

In contrast to El Leoncito, whose $T_L$ distributions are largely method-independent, the Omarkhayyam asteroid shows a substantial method-dependent spread in median $T_L$ estimates: approximately 162, 581, and 856 thousand years, respectively. One possible interpretation of these divergent 
$T_L$ estimates is that Omarkhayyam undergoes intermittent interactions with weak mean-motion resonances, such as the 20:11 resonance with Jupiter or the 9:2 resonance with Saturn. These weak resonances are not strong enough to destabilize the orbit but can induce intermittent episodes of mixed dynamical behaviour, where the trajectory alternates between more regular and more chaotic phases. As a result, different numerical schemes may capture different segments of this behaviour, leading to the observed discrepancies in $T_L$ estimates, while the orbit remains overall long-term stable.

At the same time, we note that an alternative explanation cannot be excluded. As pointed out by \citet{Tancredietal2001}, renormalization-based methods may yield less reliable Lyapunov time estimates for orbits that are close to regular or only weakly chaotic. In this context, part of the observed spread in $T_L$ for Omarkhayyam may reflect methodological limitations rather than purely dynamical effects. With the present analysis, which does not include a detailed study of the resonant angle behaviour and its correlation with the Lyapunov time, we cannot unambiguously discriminate between these two interpretations.

The spread of $T_L$, as measured by the $IQR$, is small -- about an order of magnitude smaller than the median $T_L$. For this object, the skewness of the $T_L$ disribution varies depending on the method used: both the variational method and N2-variant of the renormalization method yield positive skewness (right-skewed), whereas the other renormalization , N1-variant, yields negative skewness. Given the mixed dynamical behaviour of this object discussed above, such skewness values may be misleading and should be interpreted with caution.

Our results are in good agreement with the $T_L$ estimates reported in the AFP and AstDyS-2 databases, as well as with those obtained using the variational method by \citet{HolmanandMurray1996}. In contrast, the renormalization method applied by \citet{HolmanandMurray1996} yielded $T_L$ values an order of magnitude lower than the other estimates. All these parameters are summarized in Table~\ref{tab:MBA-LT}.

\subsubsection{Unstable orbits}
\label{Sect:oMBA_Unstable}
\begin{figure*}
	\centering	\includegraphics[width=8.00cm]{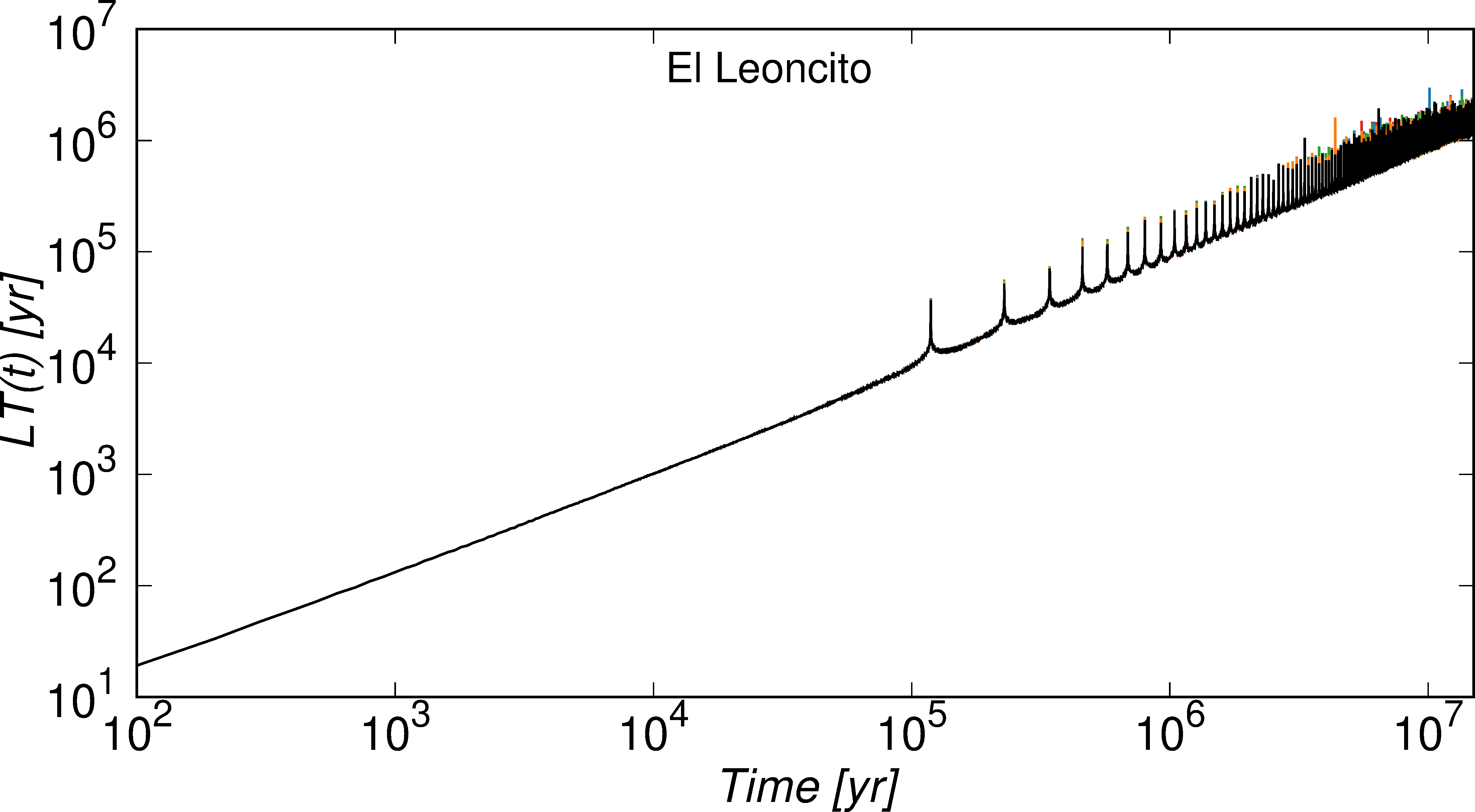}
	\includegraphics[width=8.00cm]{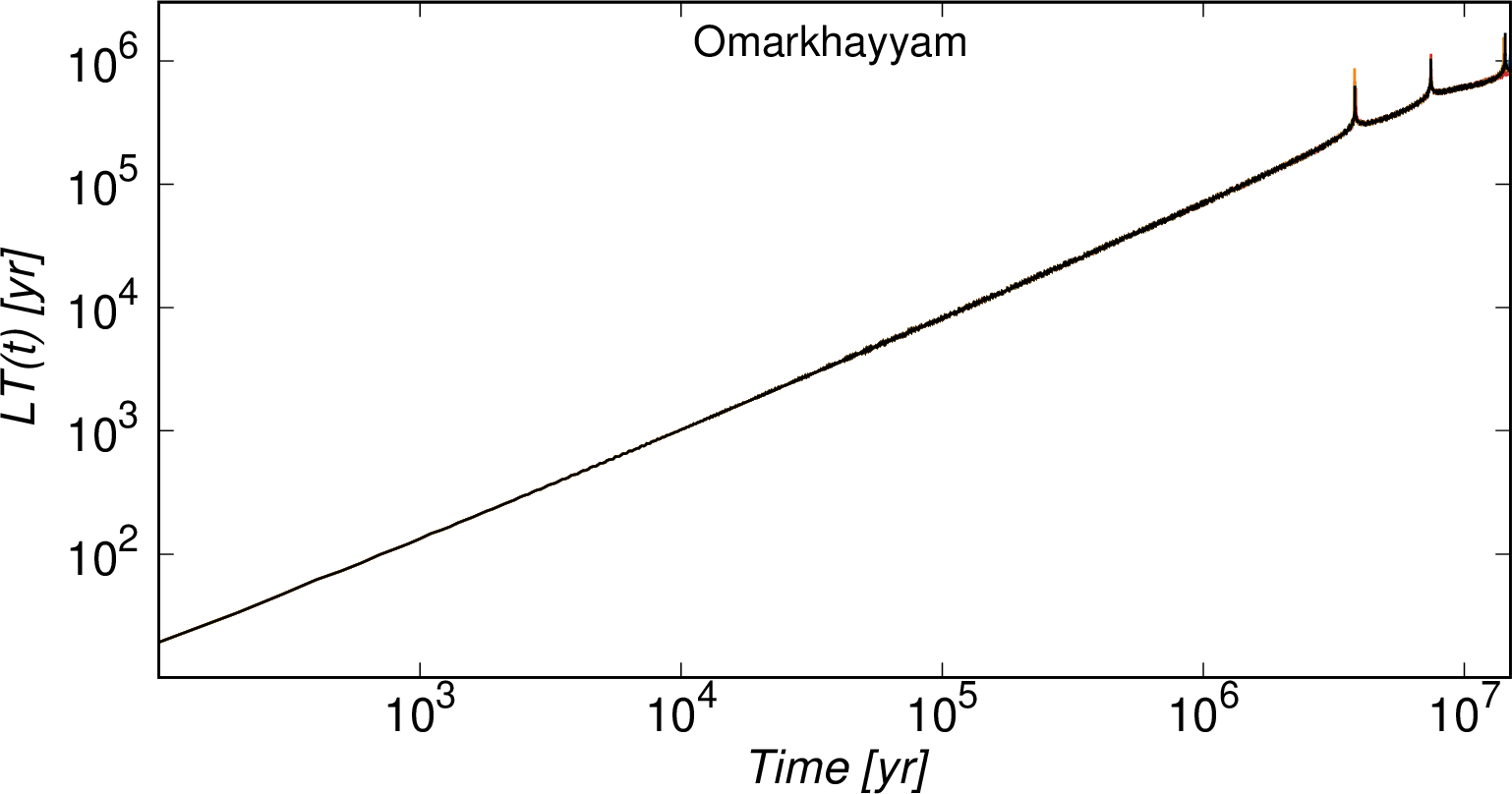}\\
	\includegraphics[width=8.00cm]{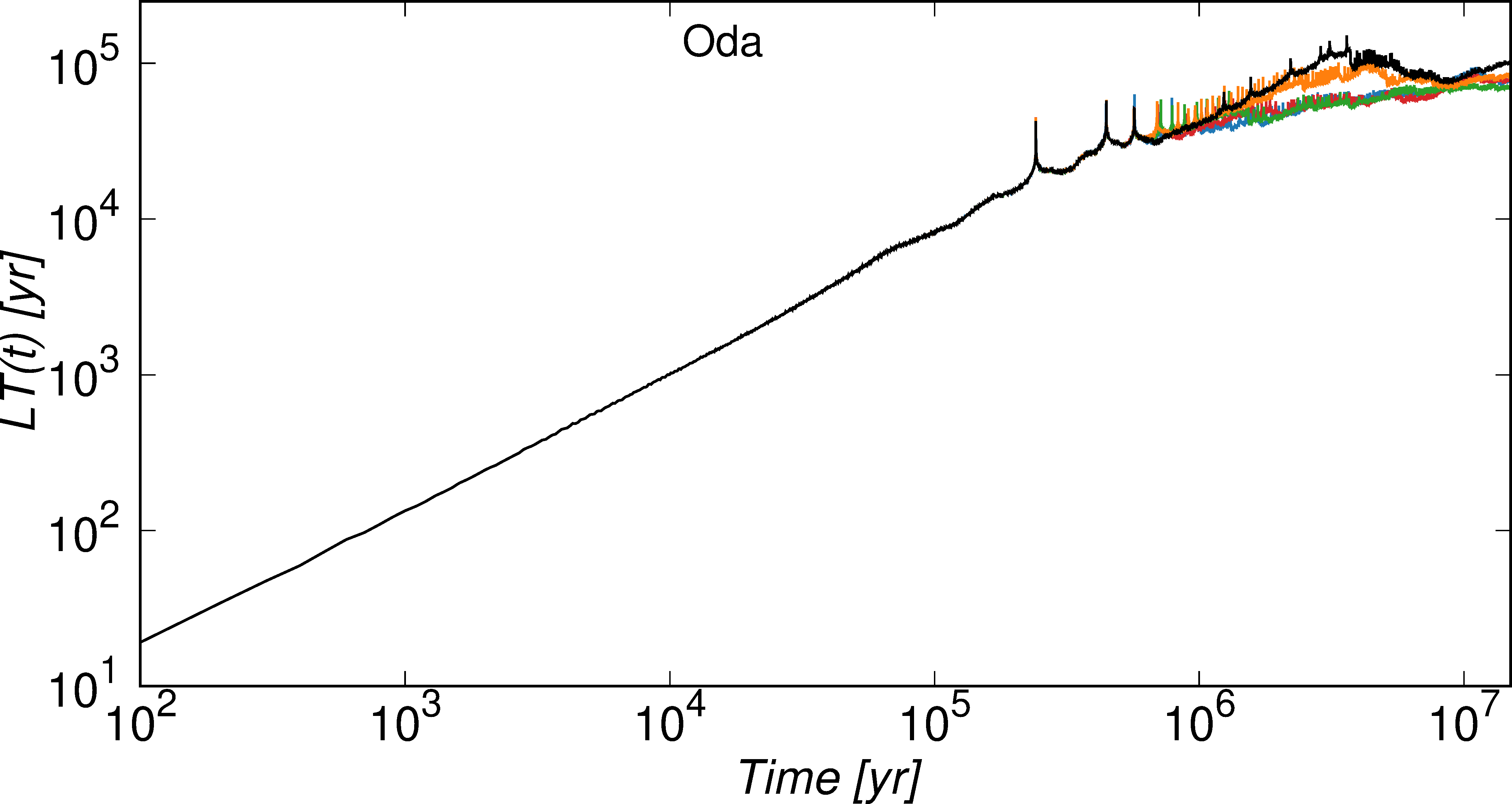}
	\includegraphics[width=8.00cm]{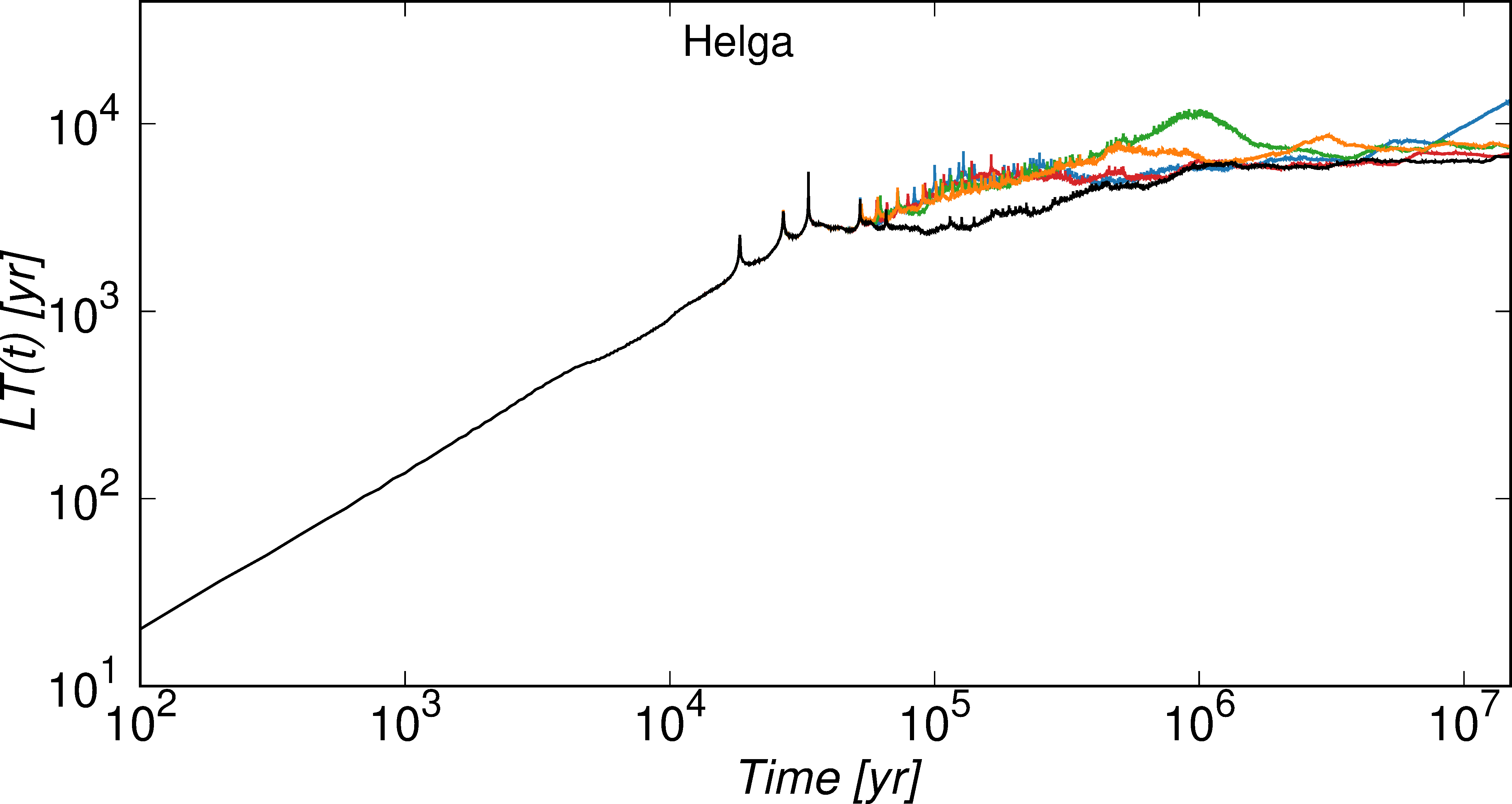}\\

\hspace{8cm} \includegraphics[width=8.00cm]{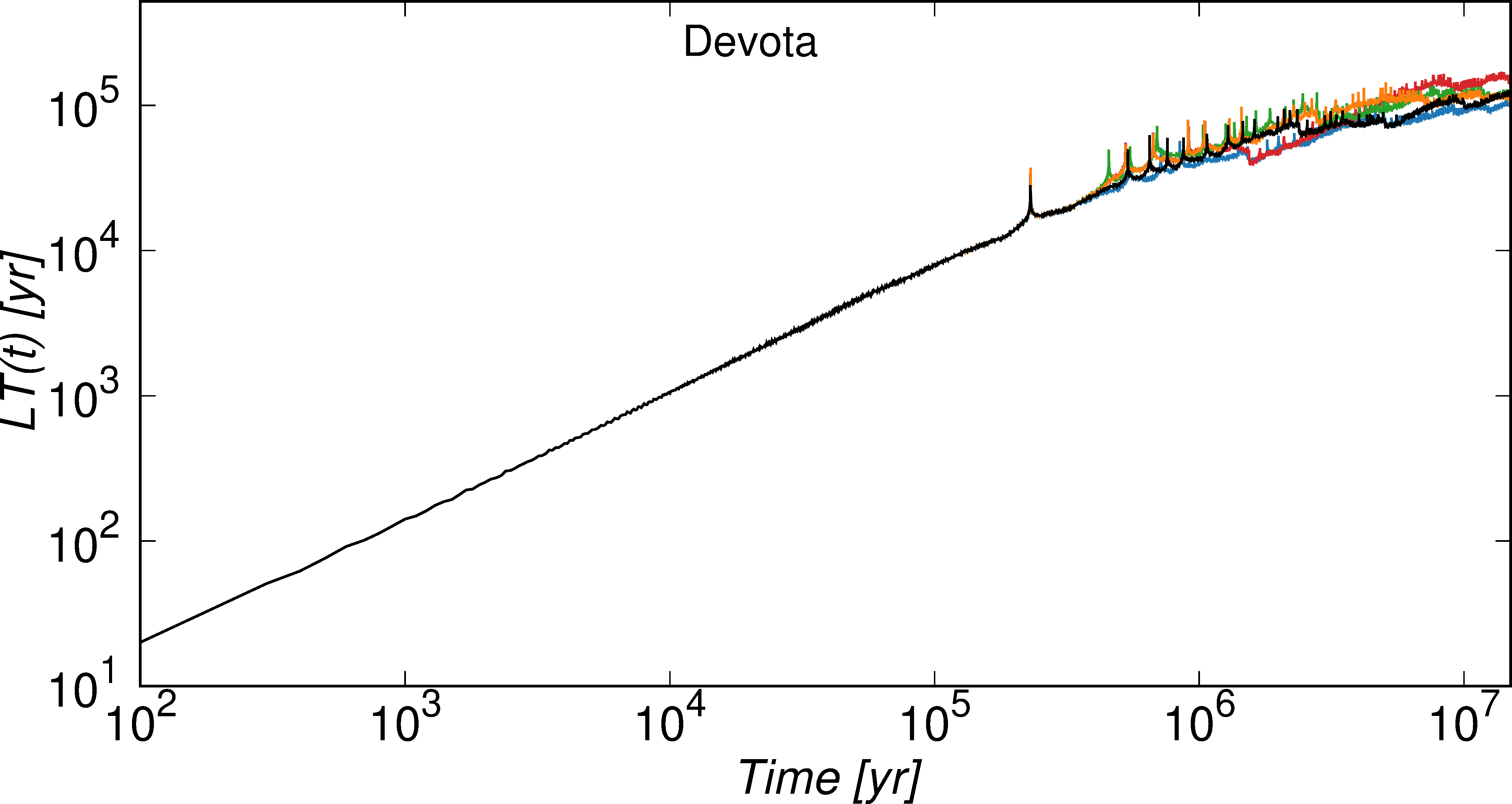}
	\caption{Time evolution of $T_L(t)$ for outer Main Belt objects, showing the nominal orbit (black line) and four randomly selected virtual objects, calculated using the variational method.}	
	\label{fig:evol_Lapunov_oMBA}
\end{figure*}

\begin{figure*}
	\centering
	\includegraphics[width=8.00cm]{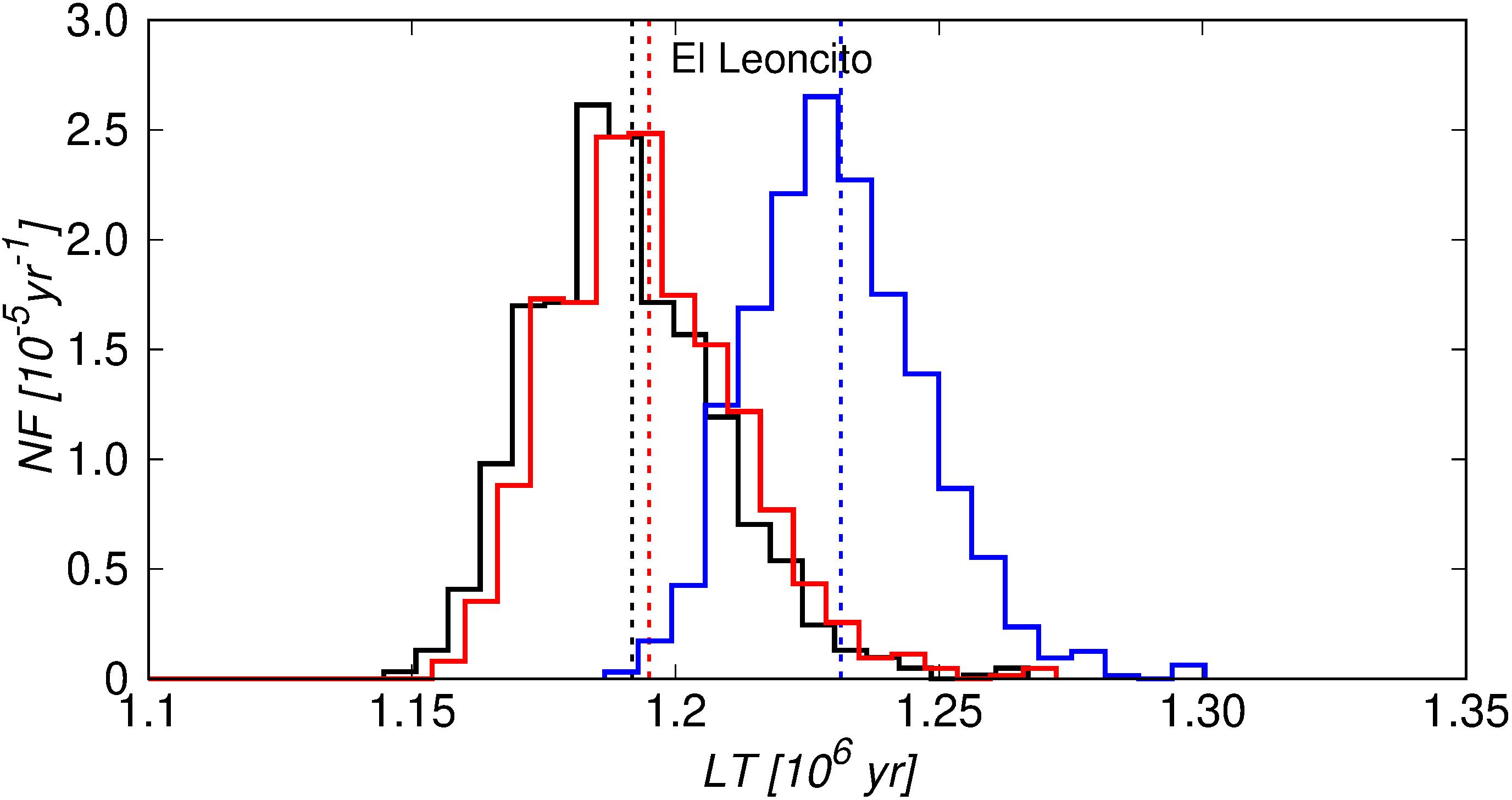}
	\includegraphics[width=8.00cm]{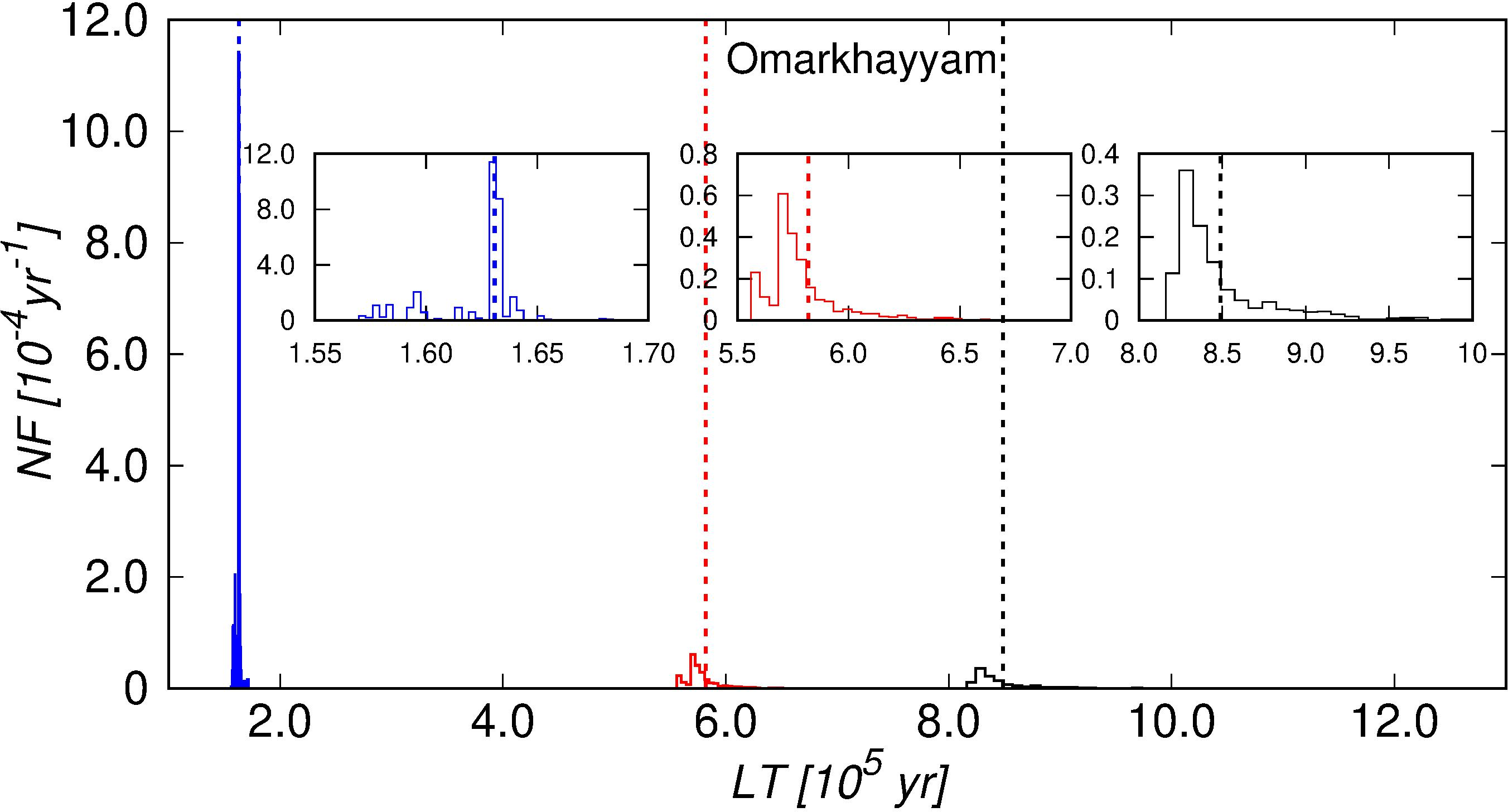}\\
	\includegraphics[width=8.00cm]{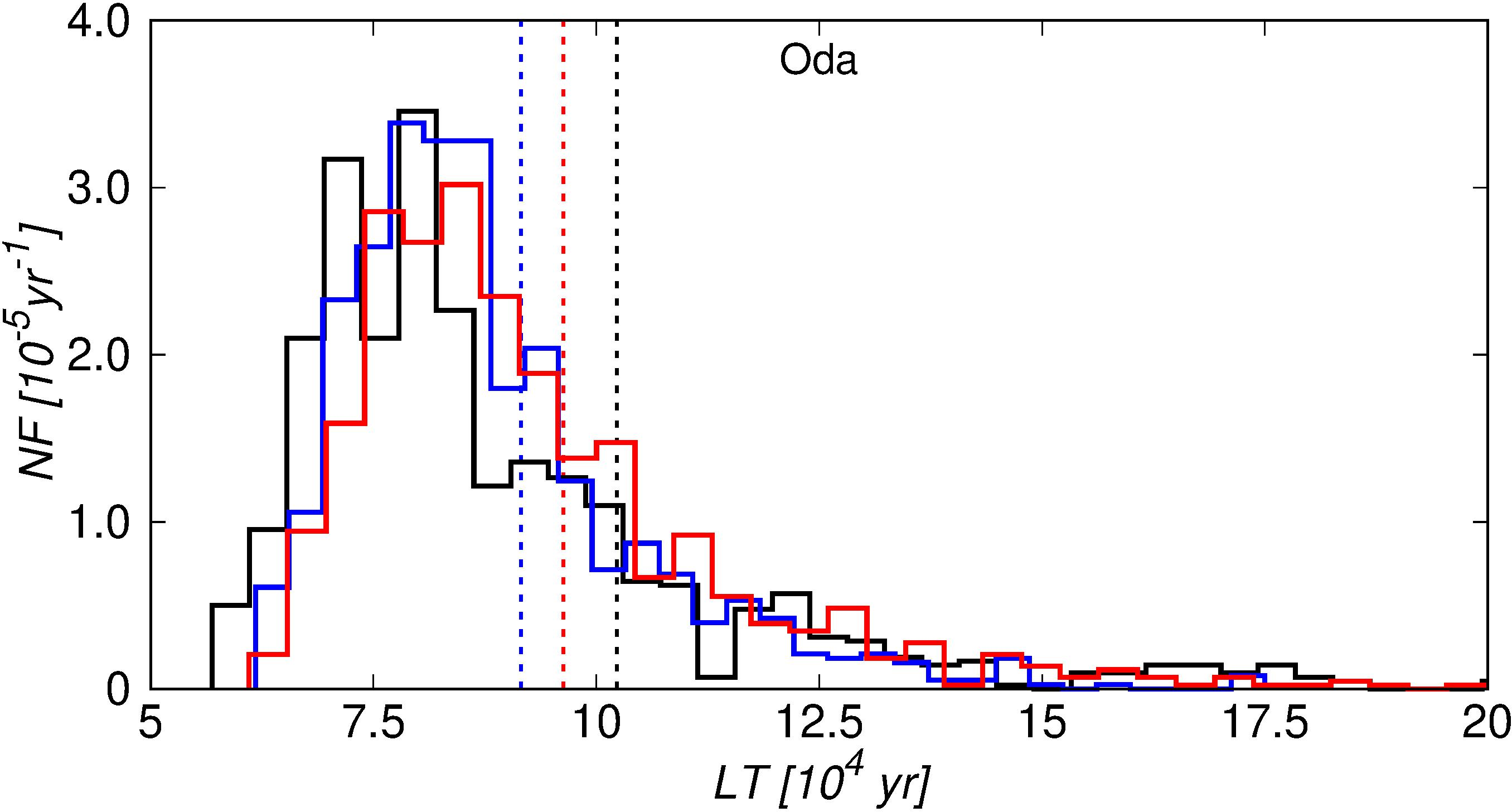}
	\includegraphics[width=8.00cm]{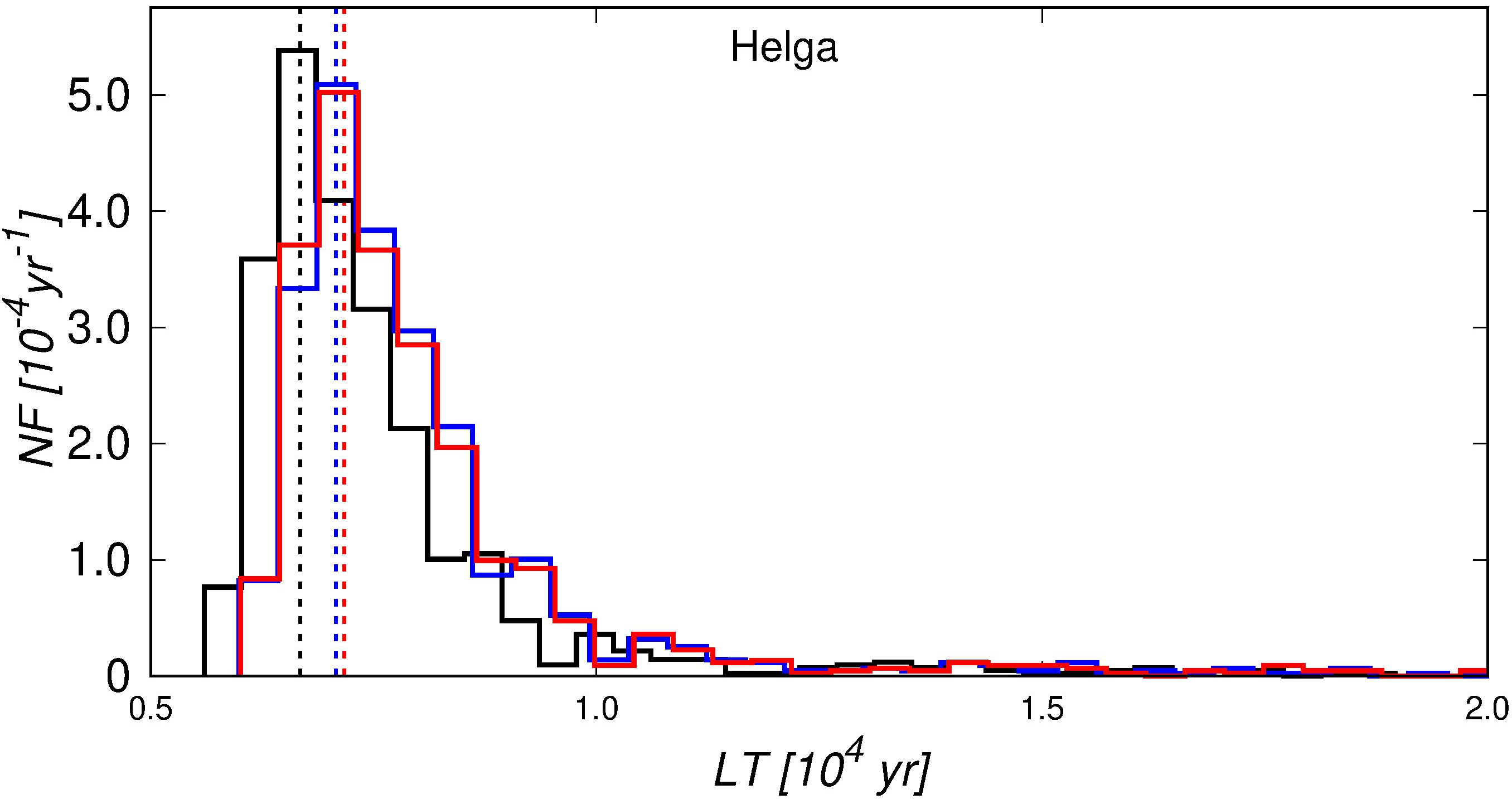}\\

\hspace{8cm}\includegraphics[width=8.00cm]{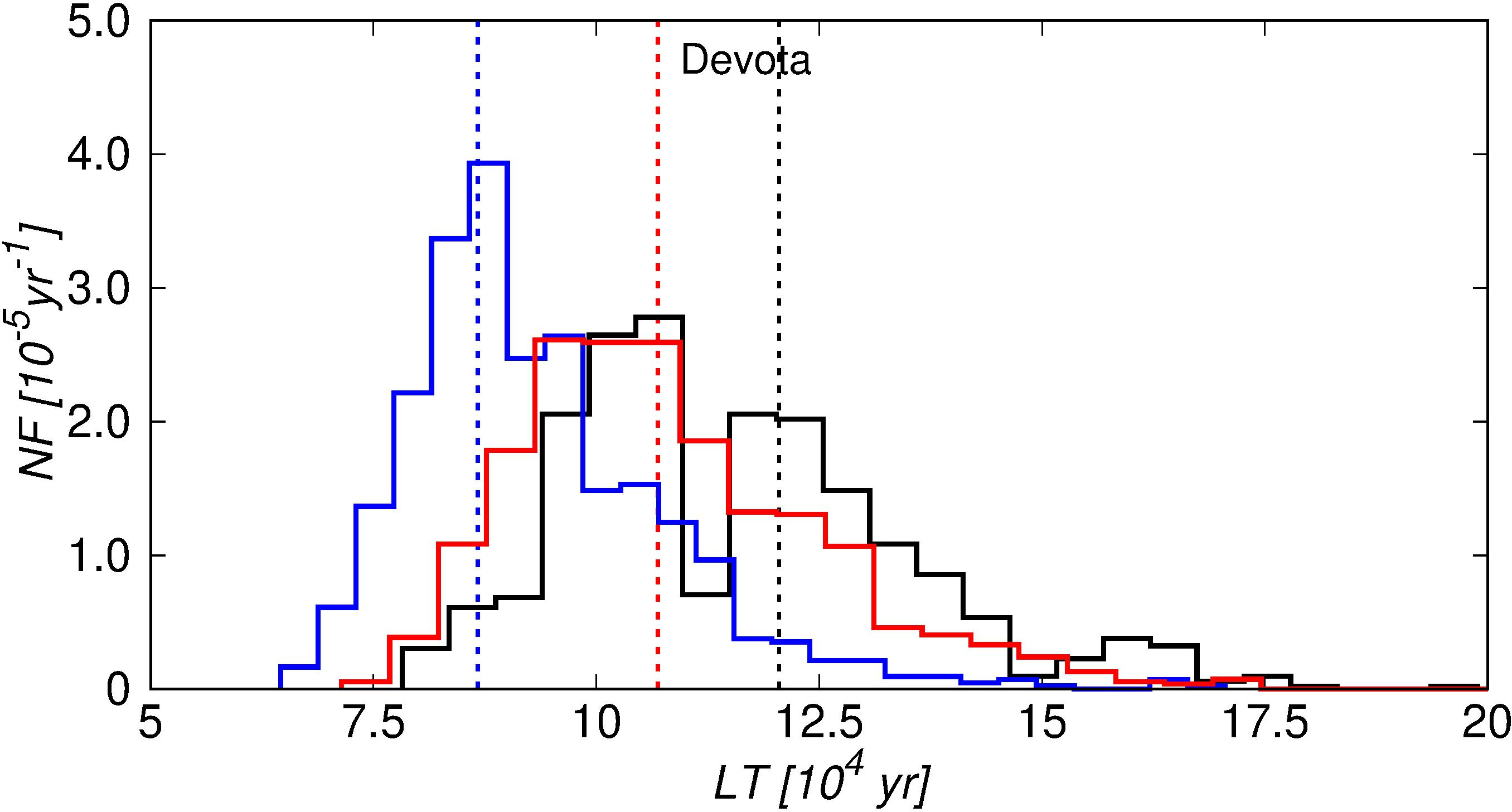}
	\caption{Statistical distribution of Lyapunov time calculated for the swarm of VAs after 15\,Myr for GR orbits representing outer MBA objects analysed in this study. The histograms show the calculation results performed by two methods: variational (black line) and neighbour trajectories method in two variants (blue and red lines). The $T_L$ for the nominal orbit is represented by the vertical dashed lines.}	\label{fig:Lapunov_oMBA}
\end{figure*}

The group of objects with unstable orbits comprises Oda, Helga, and Devota. As shown in Fig.~\ref{fig:evol_Lapunov_oMBA}, the convergence of $T_L(t)$ toward a constant value is most clearly observed for Helga. For Oda and Devota, $T_L(t)$ exhibits a tendency toward convergence; however, the 15\,Myr integration interval is likely insufficient to fully capture the asymptotic behavior. Therefore, in the case of these two objects, the $T_L$ is likely greater than our estimate. For all three objects, the methods we used yielded fairly consistent Lyapunov time estimates, both for the median $T_L$ and for the nominal orbit. The largest discrepancies are observed in the $T_L$ estimate for Devota (see Fig.~\ref{fig:Lapunov_oMBA} and Table~\ref{tab:MBA-LT}). This can be explained in a similar way as for Omarkhayyam, discussed in the previous subsection. Like Omarkhayyam, Devota lies close to the 20:11 resonance with Jupiter, the 9:2 resonance with Saturn, and additionally the 9:5 resonance with Jupiter, and may occasionally be captured in these resonances.

Our calculations indicate that Helga’s orbit is the most chaotic, with a median $T_L$ of approximately 8--9 thousand years. In contrast, the orbits of the other two asteroids are less chaotic, with median $T_L$ values on the order of 100 thousand years. Our $T_L$ estimates for these three objects are of the same order of magnitude as the results from the AFP and AstDyS-2 databases. The results obtained by \citet{HolmanandMurray1996}, particularly the $T_L$ estimates using the variational method, suggest that the orbits of Oda and Helga are stable, with $T_L$ values exceeding 1 million years. It should be noted that these calculations were performed in a different model, namely the ERTBP, as mentioned earlier in Section~\ref{Sect:oMBA_Intr}. On the other hand, their estimate of $T_L$ for Devota is approximately 4 thousand years, indicating the instability of this object. A similar value of $T_L$ (about 3.4 thousand years) was obtained by \citet{Murisonetal1994}. 

It is worth noting that the variation of our $T_L$ estimates for Helga is quite large. We also found that for some orbits, their $T_L$ is as much as 10 times the average of the whole sample. For all these objects the $T_L$ distributions are right skewed. The only exception is the histogram of this parameter for Devota calculated by the variational method, which is slightly left-skewed.

\begin{table*}
\caption{\label{tab:MBA-LT}Statistical parameters describing the distribution of Lyapunov times calculated for 1001 VAs of outer main-belt asteroids (MBAs) analysed in this study. The meaning of the listed parameters ($Med$, $IQR$, $A_Q$, $T_L^\mathrm{nom}$, $\delta_{Med}$, $\delta_{nom}$, V, N1, N2) is the same as in Table~\ref{tab:TNO-LT}.
The last four rows summarize Lyapunov time estimates reported in the literature ($T_L^\mathrm{lit}$): the first two rows present results from \citet{HolmanandMurray1996}, the third from \citet{Murisonetal1994}, and the fourth from the Asteroid Families Portal \citep{NovakovicandRadovic2019}. The last column provides data retrieved from the AstDyS-2 database \citep{KnezevicandMilani2012}. See the main text for further details. All dimensional quantities in the table are expressed in Julian years.}
\centering
\begin{tabular}{ccccccc}
\hline \hline 			 
\multicolumn{2}{c}{\backslashbox{Par/method}{Object}} & (2311) El Leoncito & (3095) Omarkhayyam  &  (1144) Oda & (522) Helga & (1328) Devota \\ 
\hline  \hline 
 & V& 1.1920$\times 10^6$ & 8.4089$\times 10^5$ & 8.3740$\times 10^4$ & 7265 & 1.1679$\times 10^5$ \\ 
$Med$ & N1& 1.2328$\times 10^6$ & 1.6325$\times 10^5$ & 8.6237$\times 10^4$ & 7743 & 9.2391$\times 10^4$ \\ 
 & N2& 1.1961$\times 10^6$ & 5.7595$\times 10^5$ & 8.9386$\times 10^4$ & 7737 & 1.0873$\times 10^5$ \\
  \hline
& V &2.2380$\times 10^4$ &2.6435$\times 10^4$ &2.4353$\times 10^4$ &1.2864$\times 10^3$ &2.5586$\times 10^4$ \\
IQR & N1 &2.1590$\times 10^4$ &1.0700$\times 10^3$ &1.7925$\times 10^4$ &1.3889$\times 10^3$ &1.7869$\times 10^4$ \\
& N2 &2.2980$\times 10^4$ &1.2060$\times 10^4$ &2.2955$\times 10^4$ &1.3921$\times 10^3$ &2.2287$\times 10^4$ \\ 
 \hline
& V &0.0857909 &0.448156 &0.249044 &0.227161 &-0.0290784 \\
$A_Q$ & N1 &0.0662344 &-0.773832 &0.16379 &0.231154 &0.218128 \\
& N2 &0.0765883 &0.473134 &0.249414 &0.245545 &0.172707 \\
 \hline

& V & $1.1918 \times 10^6$ & $8.4860 \times 10^5$ & $1.0229 \times 10^5$ & $6.6751 \times 10^3$ & $1.2047 \times 10^5$ \\
$T_L^\mathrm{nom}$ & N1 & $1.2314 \times 10^6$ & $1.6307 \times 10^5$ & $9.1531 \times 10^4$ & $7.0761 \times 10^3$ & $8.6661 \times 10^4$ \\
 & N2 & $1.1950 \times 10^6$ & $5.8175 \times 10^5$ & $9.6274 \times 10^4$ & $7.1720 \times 10^3$ & $1.0686 \times 10^5$ \\
\hline 
$\delta_{Med}$ & &3.0442$\times 10^{-2}$   &1.2866$\times 10^{0}$  &3.6424$\times 10^{-2}$& 7.9138$\times 10^{-4}$&2.3020$\times 10^{-1}$
 \\
\hline 
$\delta_{nom}$ & &3.0191$\times 10^{-2}$ &1.2907$\times 10^{0}$  &7.1844$\times 10^{-2}$ &1.3765$\times 10^{-2}$ &3.2299$\times 10^{-1}$
\\  

\hline \hline 
 & Var&5906  &$>1.13743\times 10^6$  &$>1.06748\times 10^6$ &$>1.00040\times 10^6$ &4329 \\ 
 & rescale (1) & 5930 &61565  &105163 &77885 & 4329 \\ 
$T_L^\mathrm{lit}$ & rescale (2) & 5005 &$>$58114 &$>$24906  &$>$34572  & 3404 \\  
& AFP &$6.667 \times 10^5$   & $1.00000\times 10^7$  &204100   & 6600&30900\\
&AstDyS-2 &404858 & 558659 & 69013 & 6858 & 50378
 \\ 
\hline 
\end{tabular} 
\end{table*}

\section{Orbital accuracy and its effect on Lyapunov time estimation}
\label{Sect:TNO_oMBA_Accuracy}
In this section, we discuss the possible relationship between two aspects: on the one hand, the fact that the actual orbit of an object is only approximately known (which motivates the generation of a cloud of VAs to better understand its orbital evolution, at least in a statistical sense), and on the other hand, the estimation of $T_L$ for the nominal orbit. 

In principle, the accuracy of an orbit (i.e., the width of the VAs cloud in the orbital parameters such as $a$, $e$, and $i$) does not directly affect the $T_L$ of an individual orbit, in particular the nominal one, since $T_L$ measures the local divergence of nearby trajectories and is mainly determined by the dynamical environment (mean motion/secular resonances, planetary perturbations). Nevertheless, two indirect aspects may play a significant role. 

Firstly, the $T_L$ of the nominal orbit should be understood as the stability estimate for the best-fit orbital solution derived from the available observations. This nominal solution may not perfectly coincide with the “true” orbit of the object, which would correspond to an ideal set of orbital elements. Moreover, as new observations are added, the best-fit solution may shift slightly in the orbital-element space, which in turn can lead to a different value of $T_L$ for the updated nominal orbit. For this reason, calculating $T_L$ for a whole cloud of VAs can be regarded as sampling the average stability in the neighbourhood of the nominal orbit, thereby providing a more realistic picture of the object’s dynamical behaviour.  

Secondly, the distribution of $T_L$ values across the VA population depends strongly on orbital uncertainty, measured for example by $\delta a/a$ (see Sect.~\ref{Sect:Data}). If the orbital elements of the object are subject to large uncertainties, the clones spread over a larger region of phase space\footnote{To be more precise, this spread also depends on the dynamical environment of the orbital elements in phase space. For example, two objects with orbits determined with the same level of accuracy may exhibit different rates of divergence among their clones, depending on their dynamical location.}. Some  may fall into chaotic zones, while others remain in more stable regions. This leads to a broad or even multimodal distribution of $T_L$, complicating the interpretation of the object’s long-term dynamics. An example of such behaviour is observed for 2010 HE$_{79}$, where the values of $T_L$ concentrate around three maxima. A similar situation, although less pronounced, can be observed for 2010 EL$_{139}$ (see Sect.~\ref{Sect:TNO_Unresolved}).

This second point implies that the median, or any other statistical indicator used to characterize $T_L$ of the cloud of VAs, may be subject to systematic bias in the presence of large orbital uncertainties. A wide VA set may include clones with very short $T_L$, which lowers the overall indicators and gives the impression of stronger instability. In contrast, when the orbit is well determined, the VA cloud is narrower, the distribution of $T_L$ is much more coherent, and the dynamical characterization of the object becomes more reliable. However, the investigation of a rich VA cloud will always be more conclusive than the analysis of the nominal orbit alone.
	
\section{Comparison of estimation discrepancies in Lyapunov time for TNOs and MBAs}
\label{Sect:TNO_oMBA_Comparison}

\begin{figure*}[!htbp]
\centering
\includegraphics[width=18.00cm]{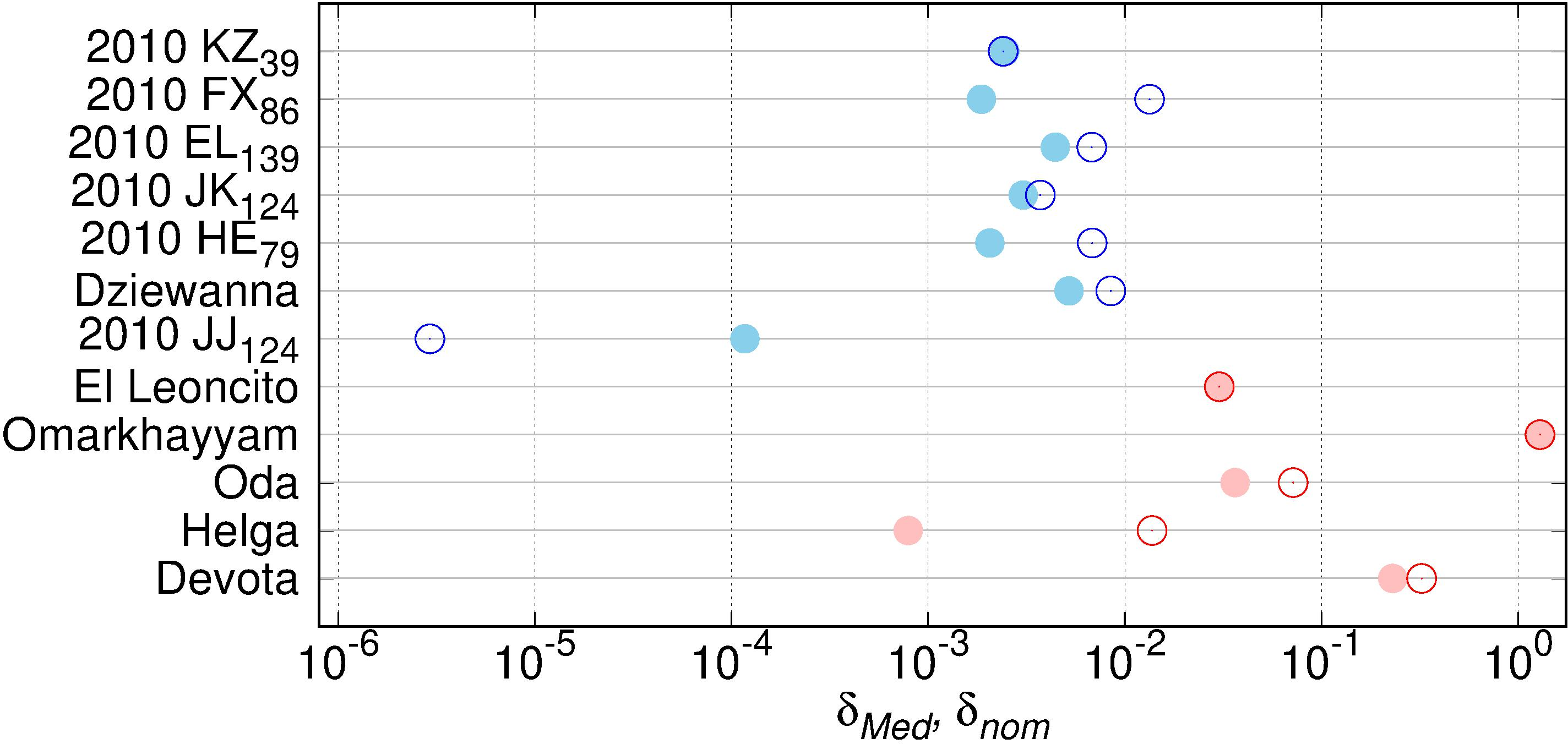}	
\caption{Comparison of the relative spread parameters $\delta_{Med}$ and $\delta_{nom}$ for individual objects from both populations. The horizontal axis shows the relative spread on a logarithmic scale, while the vertical axis lists the names of the analysed objects. Light red circles represent $\delta_{Med}$ values for Main Belt Asteroids (MBAs), while light blue circles correspond to Trans-Neptunian Objects (TNOs). Red open circles denote $\delta_{nom}$ values for MBAs and blue open circles indicate $\delta_{nom}$ values for TNOs.}
\label{fig:Delta_TNO_MBA}
\end{figure*}

The $T_L$ values presented in Sect.~\ref{Sect:Results}, obtained using various computational methods, demonstrate that this parameter can vary for a given object depending on the method applied. This concerns both the $T_L$ estimated for the nominal orbit and the indicators derived from a VA cloud, represented by the median of the $T_L$ distribution. In the majority of cases analyzed in this work, results obtained using different methods are consistent; however, for some objects, the discrepancies are substantially larger. One of the objectives of this study was to examine how the use of $T_L$ values obtained from VAs affects the consistency between different estimation methods. For this purpose, in Sect.~\ref{Sect:Stat} we defined two indicators: the relative spread of the median Lyapunov time ($\delta_{Med}$) and the corresponding spread for the nominal orbit ($\delta_{nom}$). These serve as measures of consistency across different numerical methods applied to the same object. The results of our analysis are presented in graphical form in Fig.~\ref{fig:Delta_TNO_MBA} and in numerical form in Tables~\ref{tab:TNO-LT} and~\ref{tab:MBA-LT}.

From the above-mentioned figures and tables, a general trend can be identified: the spread of $T_L$ estimates, whether measured by $\delta_{Med}$ or by $\delta_{nom}$, is typically smaller for TNOs than for oMBAs. For the analyzed TNOs, we find $\delta \lesssim 2.0 \times 10^{-2}$, while for oMBAs the corresponding values are usually $\delta \gtrsim 2.0 \times 10^{-2}$. Here, $\delta$ refers to both indicators, $\delta_{Med}$ and $\delta_{nom}$. There are, however, two noteworthy exceptions to this rule: the TNO 2010 FX$_{86}$, with $\delta_{nom} = 1.33 \times 10^{-2}$, and the oMBA Helga, with $\delta_{nom} = 7.91 \times 10^{-4}$. These cases highlight that the separation at $\delta \simeq 10^{-2}$ should not be interpreted as a physically meaningful threshold between the dynamical behaviours of TNOs and oMBAs.

It is particularly worth noting the case of asteroid Helga. Helga is a well-known resonant asteroid exhibiting stable chaotic motion, characterized by a relatively short Lyapunov time but limited variations of orbital elements such as $a$, $e$, and $i$. In this context, the very small value of $\delta_{nom}$ indicates that different numerical methods yield highly consistent estimates of $T_L$, despite the chaotic nature of the orbit. This suggests that, for this type of object, the numerical estimation of Lyapunov time is largely independent of the method used. Although our sample is too small to consider this a general diagnostic criterion, the combination of a short $T_L$ and a small $\delta_{nom}$ may serve as a characteristic sign of stable chaotic motion.

What is important to stress, however, is the broader conclusion: in general, $T_L$ estimates (for both individual nominal orbit and the median $T_L$ for the cloud of VAs) for TNOs show greater consistency across different numerical methods than those for oMBAs. This difference may be explained by the fact that TNOs, lying farther from the main planetary perturbations of Jupiter and Saturn, often evolve in dynamically simpler regimes, whereas oMBAs are more strongly affected by a dense network of mean-motion and secular resonances, which enhances the sensitivity of $T_L$ estimates to the chosen numerical method. Importantly, our statement that TNOs show greater consistency across different numerical methods should be understood as referring strictly to the high-inclination (“hot”) TNOs in our analyzed sample. The current study does not include low-inclination “cold classical” TNOs, and therefore our results cannot be generalized to the full TNO population.

For both MBAs and TNOs, the relative spread is generally smaller when calculated for the VAs ($\delta_{Med}$) than for the nominal orbit ($\delta_{nom}$). The only exception to this trend is 2010 JJ$_{124}$, where $\delta_{nom}$ is of the order of $10^{-6}$, while $\delta_{Med}$ reaches approximately $10^{-4}$. In the case of (32740) El Leoncito, (109879) Omarkhayyam, and 2010 KZ$_{39}$, the two parameters are nearly identical. This behaviour is particularly noteworthy, as it suggests that the use of a cloud of VAs leads to more consistent $T_L$ estimates across different methods than computations based solely on the nominal orbit. It could be tentatively explained as follows. When we use VAs to estimate the median $T_L$, we effectively sample the neighbourhood of the nominal orbit in phase space, which helps to smooth out method-dependent biases that may affect $T_L$ estimates when calculated for a single nominal orbit. In other words, the nominal orbit may occasionally lie in a region that is more sensitive to the chosen numerical method, leading to larger discrepancies in the resulting $T_L$ values.

In addition, a systematic comparison with the variational method indicates that, for the majority of the analysed objects, the renormalization method with the smaller initial separation (N2) yields Lyapunov-time estimates that are closer to the variational-method results than those obtained with the N1 scheme. This behaviour is observed for most of the analysed MBAs and TNOs, with only a few exceptions. Although this trend is not universal, it suggests that, in many practical applications, the choice of a tighter initial deviation vector in renormalization-based approaches may lead to improved agreement with variational Lyapunov time.

\section{Closing Summary}\label{Sect:Summary}

In this study, our main goal was to investigate and compare different approaches for estimating the Lyapunov time of small Solar System bodies and to assess their reliability for objects moving in different dynamical regimes. We applied three methods for computing the maximal Lyapunov exponent: the variational method and two implementations of the renormalization method. For each of the twelve selected objects, we calculated $T_L$ not only for the nominal orbit but also for ensembles of 1001 virtual clones, allowing us to compare single-orbit estimates with clone-based statistics. The selected targets -- five outer main-belt asteroids and seven trans-Neptunian objects -- were chosen to represent distinct dynamical classes and resonance conditions, providing a broad testing ground for the methods. A second key aim of this work was to highlight the advantages of ensemble-based $T_L$ calculations, which average over the local dynamical neighborhood of the nominal orbit. We also examined the issue of Lyapunov time convergence and its implications for the predictability of orbital stability across different dynamical regimes.

All applied methods demonstrated very good agreement in estimating Lyapunov times across the studied sample. However, a general trend observed in our results is that trans-Neptunian objects show greater consistency across different numerical methods than outer main-belt asteroids. The most pronounced differences were found for the oMB asteroids Omarkhayyam and Devota. These variations can likely be attributed to weak, intermittent interactions with mean-motion resonances (e.g., 20:11 with Jupiter or 9:2 with Saturn). As a result, different numerical schemes may capture different aspects of the orbital behaviour, leading to discrepancies in $T_L$ estimates. A possible explanation for the generally higher consistency obtained for TNOs is that their dynamical environments, although strongly perturbed by Neptune, are less densely populated with overlapping resonances than the outer main belt, where the combined influence of Jupiter and Saturn produces a more complex and sensitive resonance structure. This makes MBA orbits more prone to method-dependent variations in $T_L$ estimation.

The use of ensembles of 1001 virtual clones proved especially valuable in this context. For two TNOs, 2010 EL$_{139}$ and 2010 HE$_{79}$, the clone populations revealed a dual character: while some clones indicated stable orbital behaviour, others displayed clear signs of chaotic evolution. Such cases would have been difficult to recognize if only the nominal orbit had been considered. This highlights how ensemble-based analyses provide deeper insight: beyond quantifying orbital uncertainties, they can uncover subtle dynamical pathways that remain invisible in single-orbit studies.

Importantly, our results demonstrate that the use of VA ensembles leads to more consistent $T_L$ estimates across different methods compared to computations based solely on the nominal orbit. Median values derived from the clone populations effectively smooth out method-dependent biases and provide a more reliable characterization of dynamical stability. In this sense, our results highlight how ensemble-based $T_L$ calculations, which incorporate virtual clones to sample the local orbital neighborhood, result in a more reliable and robust characterization of orbital stability than analyses based solely on nominal orbits. This approach is particularly valuable for objects with poorly constrained orbits or those evolving in complex resonant environments.

Finally, while our study has focused on a carefully chosen but limited sample, the methodology developed here is directly applicable to larger populations of small Solar System bodies. In particular, extending this approach to systematically assess the dynamical stability of entire classes of TNOs or main-belt asteroid families could provide new insights into their long-term evolution. 

\section*{Acknowledgements}
\textbf{Paweł Wajer}: conception of the study, development of the methodology for Lyapunov time ($T_L$) analysis, final $T_L$ calculations, and drafting of the manuscript (except where noted below).
\textbf{Małgorzata Królikowska}: orbital determination for the studied objects, preparation of initial conditions for $T_L$ computations, writing of Section 3, and comments and revisions to the manuscript.
\textbf{Jakub Suchecki}: preliminary $T_L$ calculations for selected objects, which were used by PW to develop the final calculation methodology, co-writing of Section 2.
All authors: selection of the objects for analysis.

This research has made use of positional data of analyzed objects provided by the International Astronomical Union’s Minor Planet Center. 

We are grateful to the anonymous reviewer for their insightful and constructive comments, which greatly strengthened the manuscript.

\section*{Data Availability}
The data underlying this article will be shared on reasonable request to the corresponding author.



\bibliographystyle{mnras}
\bibliography{references} 



\newpage
\appendix

\section{Orbital elements of the analyzed objects}
\label{sec:orbital-elements}
\onecolumn
\begin{landscape}

\begin{table}
\caption{\label{t7} General characteristics of purely gravitational orbits of the analyzed objects taken from MPC database in 2023 May~25. Second column shows the classification according to the JPL~Small-Body Database Browser, where TNO means Trans-Neptunian Object, oMBA -- outer Main Belt Asteroid, ETC -- Encke-type Comet, and HTC  -- Halley type Comet. Abbreviations: PRE, POST, and INS in column [12] ('type of data arc') mean that perihelion passage are respectively after, before, or inside the data-arc. In each of group (TNOs, oMBC and both type of comets) objects are ordered with increasing eccentricity (column [5]) ; for more see text. }
\centering
\setlength{\tabcolsep}{5.0pt} 
\begin{tabular}{ccccccccccccc}
\hline\hline
Name & Classifi & a     & q      & e & P    & i      & Q      & data arc              & T & Epoch  & type of  \\
     & -cation  & [au]  & [au]   &   & [yr] & \degr  & [au]   &\multicolumn{3}{c}{[yyyy mm dd]}    & data arc          \\
$[1]$  & $[2]$      & $[3]$   & $[4]$    &$[5]$& $[6]$  & $[7]$    & $[8]$    & $[9]$                   &$[10]$&  $[11]$ &  $[12]$          \\
\hline\hline
\\
\multicolumn{13}{c}{\bf Trans-Neptunian Objects} \\
\\
2010 KZ39  & TNO& 45.34& 42.79& 0.0562& 305& 26.11& 47.88& 2010 05 21 -- 2022 04 29&    2103 05 16& 2022 01 21 & PRE& \\ 
2010 FX86  & TNO& 46.76& 43.81& 0.0631& 320& 25.17& 49.70& 2010 03 17 -- 2022 05 01&    2083 04 01& 2022 01 21 & PRE& \\ 
2010 EL139 & TNO& 39.16& 36.79& 0.0603& 245& 23.01& 41.52& 2010 03 12 -- 2022 05 25&    1995 02 28& 2023 02 25 &POST& \\ 
2010 JK124 & TNO& 40.37& 35.73& 0.115 & 257& 15.50& 45.02& 2010 05 11 -- 2014 06 04&    1947 04 28& 2010 05 24 &POST& \\ 
(471165) 2010 HE79           & TNO& 38.90& 32.00& 0.177& 243& 15.74& 45.80 & 1992 05 24 -- 2022 02 12& 1974 10 06& 2022 08 09 & POST&\\ 
(471143) Dziewanna           & TNO& 69.57& 32.46& 0.533& 580& 29.49& 106.7 & 2002 03 15 -- 2020 05 14& 2038 12 23& 2022 01 21 &  PRE &\\ 
2010 JJ124                   & TNO& 85.57& 23.61& 0.724& 792& 37.70& 147.5 & 2010 05 11 -- 2018 06 14& 2012 12 01& 2018 03 23 &  INS &\\ 
\\
\multicolumn{13}{c}{\bf outer Main Belt Asteroids} \\
\\
2311 El Leoncito (1974 TA1)  & oMBA& 3.657& 3.483& 0.0477& 6.99& 6.60& 3.832& 1972 05 17 -- 2023 01 09& 2022 05 24 & 2023 02 25 & 35opp\\ 
3095 Omarkhayyam (1980 RT2)  & oMBA& 3.495& 3.216& 0.0799& 6.53& 2.97& 3.774& 1974 11 12 -- 2023 03 03& 2020 09 08 & 2023 02 25 & 34opp\\  
1144 Oda (1930 BJ)           & oMBA& 3.754& 3.450& 0.0811& 7.27& 9.81& 4.059& 1930 03 29 -- 2023 03 09& 2022 03 26& 2023 02 25 &  48opp\\  
522 Helga (A904 AF)          & oMBA& 3.628& 3.314& 0.0866& 6.91& 4.42& 3.943& 1904 01 11 -- 2023 03 26& 2019 09 09 & 2023 02 25 & 66opp\\  
1328 Devota (1925 UA)        & oMBA& 3.505& 3.029& 0.136 & 6.56& 5.77& 3.980& 1925 10 24 -- 2022 12 03& 2017 05 25& 2023 02 25 &  45opp\\  
\\
\hline
\end{tabular}
\end{table}

\begin{table}
	\caption{\label{tab:starting-orbits} Orbital elements of osculating heliocentric orbits for objects described in Table~\ref{t7}, where the range of data arc used for orbit determination is shown in col.~[9]. The successive columns signify: $[1]$ -- object designation, $[2]$ -- Epoch, i.e. osculation date, 	$[3]$ -- perihelion time [TT], $[4]$ -- perihelion distance, $[5]$ 	-- eccentricity, $[6]$ -- argument of perihelion (in degrees), 	equinox 2000.0, $[7]$ -- longitude of the ascending node (in degrees), equinox 2000.0, $[8]$ -- inclination (in degrees), equinox 2000.0. Additionally, the nineth column gives semimajor axes and the tenths -- the obtained root-mean-square.}
\centering
\setlength{\tabcolsep}{5.0pt} 
\begin{tabular}{ccrrrrrrrrc}
		\hline\hline
Object     & Epoch         & $T$       & $q$            &  $e$        & $\omega$      & $\Omega$    &  $i$           & $a$           & $\delta a/a$ & RMS        \\
$[1]$      & $[2]$       & $[3]$       & $[4]$          & $[5]$       & $[6]$         & $[7]$       & $[8]$          & $[9]$         & $[10]$       & $[11]$     \\
        \hline\hline 
\\
\multicolumn{11}{c}{\bf Trans-Neptunian Objects} \\
& $[$yyyymmdd$]$&$[$yyyymmdd.dddddd$]$ & $[$au$]$       &             &  $[$\degr$]$  & $[$\degr$]$ & $[$\degr$]$    & $[$au$]$      & [$10^{-5}$]    & [arcsec]   \\
2010 KZ39  & 20230225 & 21080706.161769 &     42.59838796 &      0.05706104 &     318.655287 &      53.262445 &      26.145298 &     45.1761883062   & 9.9  &   0.21 \\
           &          &  $\pm$12.608721 & $\pm$0.00726550 & $\pm$0.00007192 &  $\pm$0.059045 &  $\pm$0.000134 &  $\pm$0.000177 & $\pm$0.0044825952   &      &        \\
2010 FX86  & 20230225 & 20850409.781165 &     43.74531089 &      0.06136793 &     357.673019 &     311.039693 &      25.173844 &     43.7453108862   & 3.2  &   0.16 \\
           &          &  $\pm$ 5.842998 & $\pm$0.00149296 & $\pm$0.00002295 &  $\pm$0.021662 &  $\pm$0.000170 &  $\pm$0.000006 & $\pm$0.0014971705   &      &        \\
2010 EL139 & 20220809 & 19951110.435430 &     36.82211130 &      0.06065390 &     208.025109 &     331.252028 &      23.015847 &     39.1997278163   & 1.7  &   0.11 \\
           &          &  $\pm$ 1.986069 & $\pm$0.00025676 & $\pm$0.00001193 &  $\pm$0.009155 &  $\pm$0.000065 &  $\pm$0.000018 & $\pm$0.0006705504   &      &        \\
2010 JK124 & 20100524 & 19470605.588590 &     35.72708481 &      0.11539191 &     221.656934 &     277.587854 &      15.504242 &     40.3874724055   & 17.8 &   0.23 \\
           &          &  $\pm$20.066453 & $\pm$0.00792059 & $\pm$0.00026187 &  $\pm$0.072828 &  $\pm$0.000705 &  $\pm$0.000270 & $\pm$0.0071894375   &      &        \\
(471165)   & 20220809 & 19741007.323407 &     32.00116459 &      0.17734480 &     281.939957 &     238.727200 &      15.738946 &     39.1564403516   &  0.6 &   0.06 \\
2010 HE79  &          &  $\pm$ 0.127469 & $\pm$0.00009164 & $\pm$0.00001484 &  $\pm$0.000860 &  $\pm$0.000031 &  $\pm$0.000101 & $\pm$0.0002364964   &      &        \\
(471143)   & 20220121 & 20381223.907312 &     32.46348210 &      0.53333380 &     284.832094 &     346.327119 &      29.489969 &     69.5646743628   & 2.3  &   0.08 \\
Dziewanna  &          &  $\pm$ 0.092109 & $\pm$0.00019833 & $\pm$0.00000975 &  $\pm$0.000816 &  $\pm$0.000030 &  $\pm$0.000013 & $\pm$0.0015846726   &      &        \\
2010 JJ124 & 20200531 & 20121224.813590 &     23.63610289 &      0.72395394 &     339.787046 &     268.808845 &      37.690709 &     85.6237642193   & 23.1 &   0.19 \\
           &          &  $\pm$ 0.099883 & $\pm$0.00028099 & $\pm$0.00007192 &  $\pm$0.001124 &  $\pm$0.000050 &  $\pm$0.000070 & $\pm$0.0198030009   &      &        \\
\\
\multicolumn{11}{c}{\bf outer Main Belt Asteroids} \\
& $[$yyyymmdd$]$&$[$yyyymmdd.dddddd$]$ & $[$au$]$       &             &  $[$\degr$]$  & $[$\degr$]$ & $[$\degr$]$    & $[$au$]$      & [$10^{-9}$]   & [arcsec]   \\
(2311)       & 20230225 & 20220524.246122 &      3.482886733 &      0.047740117 &     181.421627 &     156.542980 &       6.604165 &      3.6574960192  & 2.2  &  0.33 \\
 El Leoncito &          &   $\pm$0.000126 & $\pm$0.000000058 & $\pm$0.000000015 &  $\pm$0.000023 &  $\pm$0.000014 &  $\pm$0.000018 & $\pm$0.0000000081  &      &       \\
(3095)       & 20230225 & 20200908.743068 &      3.215573032 &      0.079927629 &     116.968463 &     292.432112 &       2.971597 &      3.4949131543  & 2.0  &  0.35 \\
 Omarkhayyam &          &   $\pm$0.000080 & $\pm$0.000000059 & $\pm$0.000000017 &  $\pm$0.000037 &  $\pm$0.000035 &  $\pm$0.000002 & $\pm$0.0000000071  &      &       \\
(1144)       & 20230225 & 20220326.329930 &      3.449770953 &      0.081092019 &     216.935151 &     156.126837 &       9.805354 &      3.7542071950  & 1.6  &  0.27 \\
 Oda         &          &   $\pm$0.000054 & $\pm$0.000000052 & $\pm$0.000000014 &  $\pm$0.000011 &  $\pm$0.000008 &  $\pm$0.000001 & $\pm$0.0000000059  &      &       \\
(522)        & 20230225 & 20260808.164285 &      3.314031129 &      0.086638582 &     248.994979 &     116.558659 &       4.418908 &      3.6283896666  & 1.1  &  0.27 \\
 Helga       &          &   $\pm$0.000052 & $\pm$0.000000045 & $\pm$0.000000012 &  $\pm$0.000018 &  $\pm$0.000016 &  $\pm$0.000001 & $\pm$0.0000000040  &      &       \\
(1328)       & 20230225 & 20231217.190674 &      3.028918310 &      0.135730870 &     173.690341 &     222.649050 &       5.767724 &      3.5046008298  & 1.3  &  0.42 \\
 Devota      &          &   $\pm$0.000056 & $\pm$0.000000056 & $\pm$0.000000016 &  $\pm$0.000023 &  $\pm$0.000021 &  $\pm$0.000002 & $\pm$0.0000000047  &      &       \\
\\  \hline
\end{tabular}
\end{table}

\end{landscape}

\twocolumn


\bsp	
\label{lastpage}
\end{document}